\documentclass{aa}

\usepackage{natbib}
\bibpunct{(}{)}{;}{a}{}{,} 
\usepackage{graphicx}
\usepackage{txfonts}
\usepackage{lipsum}
\usepackage{subcaption} 
\usepackage{lscape}
\usepackage{placeins}
\usepackage{xcolor}

\begin{document}

\title{A systematic study of the quiescent and outburst properties of X-ray-bright young stellar objects using XMM-Newton}

\author{Koki Sakuta\inst{1}\fnmsep\thanks{Corresponding author: sakuta\_k@u.phys.nagoya-u.ac.jp}
    \and Ikuyuki Mitsuishi\inst{1}
    \and Seiya Sasamata\inst{1}
    \and Daigo Nonaka\inst{1}
    \and Mai Yamashita\inst{2}
    \and Shin Toriumi\inst{3}
    \and Takato Tokuno\inst{4}
    \and Yohko Tsuboi\inst{5}
    \and Hiroshi Kobayashi\inst{1}
    }

\institute{Graduate School of Science, Tokai National Higher Education and Research System, Nagoya University, Furo-cho, Chikusa-ku, Nagoya, Aichi, Japan 464-8602
\and Interdisciplinary Faculty of Science and Engineering Department of Applied Physics, Shimane University, 1060 Nishikawatsu-cho, Matsue, Shimane 690-8504
\and Institute of Space and Astronautical Science, Japan Aerospace Exploration Agency, 3-1-1 Yoshinodai, Chuo-ku, Sagamihara, Kanagawa, Japan 252-5210
\and Department of Astronomy, School of Science, The University of Tokyo, 7-3-1 Hongo, Bunkyo-ku, Tokyo, Japan 113-0033
\and Faculty of Science and Engineering, Chuo University, 1-13-27 Kasuga, Bunkyo-ku, Tokyo, Japan 112-8551
}

\titlerunning{A systematic study of the quiescent and outburst properties of X-ray-bright YSOs using XMM-Newton}
\authorrunning{Sakuta et al.}
\date{Received July 30, 2026}

\abstract
{
Young stellar objects (YSOs) exhibit strong X-ray emission. This is widely attributed to magnetic reconnection and magnetospheric accretion. Owing primarily to limited photon statistics, observational tests of these mechanisms have often relied on simplified analyses, however. This leaves room for more precise constraints on their emission processes.
}
{
We derive the X-ray properties of X-ray-bright YSOs selected from multiple star clusters and investigate their emission mechanisms through an approach from multiple perspectives based on a detailed timing and spectroscopic analysis.
}
{
We performed a systematic search of the XMM-Newton archival observations accumulated over the past two decades and constructed a sample of 51 X-ray-bright YSOs suitable for detailed timing and spectroscopic analyses. The timing analyses were performed to identify and explicitly separate quiescent and outburst phases. 
For each phase, we performed spectroscopic analyses using multi-temperature plasma models and examined the derived physical properties through observational diagnostics and comparisons with theoretical expectations for magnetic reconnection and magnetospheric accretion.
}
{
Multi-temperature plasma structures were detected in quiescent and outburst emission. 
The relations of the timing and spectral parameters are broadly consistent with a magnetic reconnection scenario. 
The comparison with Gaia DR3 stellar parameters suggests that magnetospheric accretion alone cannot explain most of the fitted X-ray temperatures, although the coolest model components with $kT \lesssim 0.3$ keV can include a contribution from an accretion shock. 
After we separated these low-temperature components, the remaining quiescent components were clearly positively correlated in the $EM$--$kT$ plane, with magnetic loop lengths distributed around $10^{12}$ cm.
The detection of a Neupert effect, the temporal evolution of the temperature and emission measure, and the relation of the Rossby number and X-ray activity also support scenarios driven by magnetic reconnection.
The positive correlations of the parameters in the quiescent and outburst phases suggest that the X-ray emission in both phases is closely linked to magnetic reconnection.
}
{}


\maketitle
\nolinenumbers 

\section{Introduction}

Young stellar objects (YSOs) are stars in the stage before they evolve to main-sequence stars. This stage is considered key to determining the initial condition of the stellar evolution.
YSOs are in general very active in X-rays. Their emission arises from high-temperature plasma \citep[e.g.,][]{1983ApJ...269..182M, 1999ARA&A..37..363F}.
The high-energy emission from YSOs is known to heat and ionize the gas in the protoplanetary disk, thereby affecting the disk evolution and the formation of the planetary system through photoevaporation \citep[e.g.,][]{2010EAS....41..181N, 2024ApJ...976...25W}. 
Studies of YSO activities can contribute to the understanding of the heating production of calcium-aluminum--rich inclusions (CAIs) in meteorites in the young Solar System \citep{2001ApJ...548.1029S}. 
For this reason, understanding the properties and mechanism of the X-ray emission of YSOs can provide a crucial clue not only to stellar evolution, but also to the planetary system formation process. 
For the X-ray emission mechanism of YSOs, two promising scenarios have been proposed. 
One scenario is a magnetic reconnection model, where the magnetic field lines amplified by dynamo processes in the convection layer trigger reconnection inside a large-scale magnetic loop that emits X-rays, in the same way as in the Sun and in late-type main-sequence stars \citep[e.g.,][]{1999ApJ...526L..49S, 2010ARA&A..48..241B}.
The other scenario is a magnetospheric accretion model, where the accreting matter that falls from the protoplanetary disk onto the stellar surface along the magnetic field lines forms accretion shocks in the vicinity of the stellar surface, and the shock-heated high-temperature plasma emits X-rays \citep[e.g.,][]{1998ApJ...509..802C, 2016ARA&A..54..135H}.
In the magnetic reconnection model, outbursts are characterized by exponential rise and decay phases \citep{2008ApJ...688..418G}, and theoretical studies predict correlations among the rise time, decay time, plasma temperature, emission measure, magnetic loop length, and magnetic field strength \citep{1999ApJ...526L..49S, 2002ApJ...577..422S, 2003PASJ...55..653I}. 
Such expected correlations have been applied in observational studies of YSO outbursts to estimate physical properties, such as magnetic field strengths and magnetic loop sizes \citep[e.g.,][]{2016PASJ...68...90T, 2003PASJ...55..653I}.
For outbursts and quiescent emissions, extensions of solar coronal models have been used to investigate the physical conditions of stellar coronae based on observed plasma temperatures and emission measures \citep[e.g.,][]{2002ApJ...577..422S, 2020ApJ...901...70T, 2026ApJ..1001...18I}.
Moreover, characteristic features of reconnection-driven solar flares, such as the Neupert effect \citep[e.g.,][]{1968ApJ...153L..59N, 2010JApA...31..155J} and temporal offsets between temperature and emission measure evolution \citep[e.g.,][]{2007A&A...471..271R, 2013enss.confE..39B}, have also been reported in YSOs \citep[e.g.,][]{2007A&A...468..379A, 2019ApJ...882...72V, 2005ApJS..160..469F, 2007A&A...468..485F, 2008ApJ...688..418G}.
In addition, the Rossby number, a diagnostic of the stellar dynamo efficiency, correlates with the X-ray activity in late-type stars \citep{2011ApJ...743...48W}, and a similar tendency was reported for YSOs \citep{2016A&A...589A.113A}, suggesting that low-mass YSOs reside in a regime of strong dynamo-driven magnetic activity.
In contrast, the magnetospheric accretion model predicts plasma temperatures and X-ray luminosities regulated by stellar mass, radius, and mass-accretion rate \citep{1998ApJ...509..802C}.
Observationally, a positive correlation between stellar mass and X-ray luminosity has been reported for YSOs \citep{2007A&A...468..425T, 2023MNRAS.521.2427R}, and the accretion process is expected to affect their X-ray emission \citep{2016A&A...587A..81B}.
As outlined above, extensive studies have examined the X-ray emission mechanisms in YSOs through theoretical expectations and observational signatures, but the conclusions of these investigations have not been systematically examined from multiple perspectives simultaneously. This leaves room for a further discussion of the observational constraints.

Although precise timing and spectral analyses are required to understand the X-ray emission mechanism of YSOs, there remains room for a methodological improvement in previous studies. In many cases, quiescent and outburst phases were not explicitly separated, and/or single-temperature models were adopted even when intrinsically multi-thermal plasma structures were expected \citep[e.g.,][]{2003PASJ...55..653I, 2005ApJS..160..319G, 2016A&A...589A.113A, 2019ApJ...882...72V}. 
These issues are particularly critical for YSOs located in star-forming regions, where strong and spatially variable absorption, often exceeding $N_{\rm H} \sim 10^{22}\,\mathrm{cm^{-2}}$, arises not only from circumstellar material, but also from the surrounding interstellar medium. Under such conditions, low-temperature plasma components can be overlooked even when they possess large emission measures. This leads to significantly incorrect estimates of emission-measure-weighted temperatures and total emission measures. 
\cite{2003PASJ...55..653I} performed a systematic time-resolved analysis and fitted most sources with single-temperature models, and the elemental abundances or temperatures were often fixed because the photon statistics were limited. Notably, the only source analyzed with a two-temperature model revealed a cool component with a large emission measure. These issues fundamentally stem from insufficient photon statistics. 
To overcome them, we focus on X-ray-bright YSOs with high photon statistics in multiple star clusters and perform detailed timing and multi-temperature spectroscopic analyses. By systematically separating quiescent and outburst phases and applying multi-temperature plasma models, we quantitatively characterize their temporal and spectral properties and discuss their emission mechanisms from multiple perspectives. 
Furthermore, the intrinsic X-ray properties of these X-ray-bright YSOs provide observational upper limits on their environmental effect, potentially offering insight into feedback processes relevant to disk evolution.
This paper is organized as follows. In section~\ref{sec:sample_select} we describe the selection of our sample and in section~\ref{sec:analysis_and_result} our analysis method and results. In section~\ref{sec:discussion} we discuss our results in detail before we conclude in section~\ref{sec:conclusions}.

\section{Sample selection}\label{sec:sample_select}

We performed a systematic and comprehensive search of archival observations from the X-ray Multi-Mirror Mission (XMM-Newton) using the European Photon Imaging Camera pn detector (EPIC/pn) in order to construct a large sample of X-ray-bright YSOs.
Our sample was selected from the Fourth XMM-Newton Serendipitous Source Catalog, Fourteenth Data Release (4XMM-DR14; \citealt{2020A&A...641A.136W}), which is one of the largest X-ray source catalogs currently available, containing more than one million sources detected in over 10,000 observations obtained since 2000.
To identify YSO candidates within the 4XMM-DR14 catalog, we used the Milky Way Star Clusters (MWSC; \citealt{2013A&A...558A..53K}) catalog, which compiles basic physical properties for more than 3,000 star clusters in the Milky Way, such as age, distance, and celestial coordinates. These parameters were derived from astrometric and photometric information listed in the Position and Proper Motion Extended-L (PPMXL; \citealt{Roeser_2010}) and Two Micron All Sky Survey (2MASS; \citealt{Skrutskie_2006}) catalogs.
We first selected star clusters that satisfy distance and age criteria appropriate for extracting YSOs. 
We further restricted the clusters to those within 500~pc to minimize source confusion and contamination from nearby sources.
We further limited the sample to clusters younger than 10~Myr, corresponding to the typical timescale over which star formation remains active \citep{2000ApJ...540..255P}.
For star clusters that satisfied the distance and age criteria, we searched the XMM-Newton archival data released up to November 16, 2022, covering nearly two decades of observations.
X-ray counterparts of cluster member stars were identified based on positional coincidence within 10~arcsec between the X-ray source positions listed in the 4XMM-DR14 catalog and the stellar positions reported in previous studies of the corresponding clusters or individual stars, which are summarized below.
To enable high-quality timing and spectroscopic analyses, we further restricted the sample to sources with 2,500 EPIC/pn counts at least, as reported in the 4XMM-DR14 catalog, thereby securing sufficiently high photon statistics for a detailed temporal characterization and robust multi-temperature spectral modeling. 
When multiple observations satisfied this requirement for a given source, we selected the observation with the highest photon statistics.
Finally, to minimize source confusion and contamination, we excluded sources for which other X-ray sources were detected within 30~arcsec of the target position, taking into account the angular resolution of XMM-Newton.

In consequence, we selected as our sample 51 X-ray-bright YSOs from five star clusters of Chamaeleon~I, NGC~1976, NGC~1333, Mamajek~1, and ASCC~16. 
The ages of the selected star clusters are 1, 1, 6, 10, and 10~Myr and the distances are 380, 458, 324, 112, and 397~pc, respectively, as listed in the MWSC catalog.
The celestial coordinates of the sources in our sample belonging to NGC~1976, Mamajek~1, and ASCC~16 were retrieved from the WEBDA database\footnote{\texttt{https://webda.physics.muni.cz}} \citep{2008CoSka..38..435P}, which compiles stellar observational data for star clusters in the Milky Way and the Small Magellanic Cloud. 
For sources in our sample belonging to Chamaeleon~I and NGC~1333, we adopted the positions reported by \citet{2021AA...646A..46G} and \citet{2015AJ....150...17R}, respectively.
We note that although NGC~1976 is a star-forming region in Orion and was observed in the \textit{Chandra} Orion Ultradeep Project (COUP; \citealt{2005ApJS..160..319G}), all sources in our sample identified as members of NGC~1976 are located outside the COUP field of view.
Although XMM-Newton observations covering the COUP field are available, these regions are crowded with multiple X-ray sources within $\sim30$~arcsec, making it difficult to resolve individual sources with XMM-Newton.
Table~\ref{table:analyzed_star} summarizes our sample and lists basic information for each source, including the host star cluster, stellar identifier, 4XMM-DR14 source ID, celestial coordinates, EPIC/pn photon counts, and the corresponding observation ID.
Fundamental stellar parameters such as effective temperatures, bolometric luminosities, stellar masses, radii, and distances were obtained from Gaia Data Release~3 (Gaia DR3; \citealt{2021A&A...649A...1G}) when available, although this information is not available for all sources (Table~\ref{table:gaia}). 
When no distance is provided in Gaia DR3 or the distances listed in Gaia DR3 differ significantly from the cluster distances, we adopted values from \citet{2021AJ....161..147B}. When neither was available, we used the distance to the host star cluster, with the associated uncertainty taken as the cluster radius listed in the MWSC catalog.
Although their uncertainties are not explicitly given in the MWSC catalog, they agree with Gaia-based distances to within 10\% for the clusters used in this study, NGC~1976, NGC~1333, and Mamajek~1 \citep{2025ARep...69..157V, 2018ApJ...869...83Z, 2021MNRAS.504..356D}.
Since the statistical uncertainties dominate this difference, we did not consider them in our results. 
We note that UCAC4~427-010050 might not be a member of the cluster because its distance differs significantly from the cluster distance.
Most sources in our sample with available fundamental parameters are low-mass stars, with the exception of HD~36959 and HD~36960, which are classified as a high-mass star and a high-mass candidate, respectively.
We also assessed the binarity of the sample based on previous studies \citep{2002A&A...384..180F, 2004A&A...424..727P, 2019A&A...623A..72K, 2021MNRAS.506.2269E}. 
Hen~3-545 and V1878~Ori are classified as single stars, whereas HD~36959, HD~36960, and HD~287848 are identified as members of binary systems. 

\section{Analysis and results}\label{sec:analysis_and_result}
\subsection{Methods}
\subsubsection{Imaging analysis}

We defined the regions to extract light curves and spectra of the target sources as follows. 
Following the method described in \citet{2020ApJ...901...70T}, we first extracted source events from a circular region with a radius of 30~arcsec and background events from an annular region with inner and outer radii of 45~arcsec and 60~arcsec, respectively, and created redistribution matrix files and ancillary response files.
When nearby X-ray sources were present, we defined contamination regions around them as their source extents and positions listed in the 4XMM-DR14 catalog, with an additional outer rim of 30~arcsec, and excluded these regions from the source and background regions defined above.
For extended nearby X-ray sources, however, when the contaminating photons were estimated to be fewer than 10\% of the total incoming photons from the target source, we did not exclude the contamination regions from the corresponding source and background regions in our analysis to maximize the photon statistics. 
In some limited cases when a presumably single source was cataloged as multiple X-ray sources because its position lay on or near a chip gap in the EPIC/pn, we treated it as a single source.

\subsubsection{Timing analysis}

To derive detailed X-ray properties of our sample and place reliable constraints on the X-ray emission mechanisms, it was necessary to treat quiescent and outburst phases separately because the X-ray properties differ between the two phases and the dominant physical processes might differ as well.
We therefore identified outburst intervals in the light curves, analyzed the data for each phase independently, and determined the rise and decay times of the detected events.
For each source, we extracted background-subtracted 1000 s binned light curves for the 0.4--5 keV band, where the detector sensitivity was taken into account \citep{2004A&A...414..767K}. 
We searched the light curves for outbursts using the criterion of \citet{2003PASJ...55..653I}. The test was applied iteratively to each time bin: bins satisfying the criterion were identified as outbursts and were excluded, the mean count rate was recomputed from the remaining bins, and the criterion was reapplied until no further outbursts were found. This procedure minimizes the risk that small-amplitude outbursts are obscured by large events that bias the mean count rate.
The criterion equation alone is prone to false positives, perhaps due to fluctuations of the background. 
Thus, we adopted stricter criteria to define an outburst phase, requiring either (i) three consecutive bins to satisfy the criterion, (ii) two consecutive bins for which the outburst model described below was not rejected, or (iii) a single bin whose lower flux limit exceeded the mean flux of the three preceding bins by a factor of two at least.

For each outburst, we fit the light curve with a model consisting of a constant component for the quiescent emission and one or more exponentially rising and decaying components \citep[e.g.,][]{2008ApJ...688..418G, 2024ApJ...976...25W} via $\chi^2$ minimization to derive the rise and decay times.
In the fitting, we treated all model parameters as free parameters, including the normalization of the constant flux component, and for each outburst component, the peak count rate, the peak time $t_p$, the rise time $\tau_r$, and the decay time $\tau_d$. The statistical errors were estimated from the standard deviations derived from the covariance matrix.
For periods exhibiting multiple flux enhancements, we added additional outburst components to the light-curve model and assessed the statistical significance of the fit improvement using F-tests; cases significant at the 99\% confidence level were classified as separate outbursts. The same procedure was applied to tentative candidates identified by only two consecutive bins (condition ii), allowing us to capture short-duration low-amplitude outbursts.
Based on the fitted outburst model, we defined intervals in which the outburst component dominated as outburst periods and the remaining intervals as quiescent periods. As exceptions, when the outburst excess was too low for the outburst component to dominate or when the resulting outburst duration was shorter than 1500 s, we instead defined the outburst period as $[t_p - \tau_r : t_p + \tau_d]$ to ensure sufficient photon statistics for the spectral analysis. Conversely, when this interval accounted for more than 80\% of the observation, we regarded the entire observation as quiescent, interpreting the variation as long-term rather than an outburst.

\subsubsection{Spectral analysis}

We estimated the temperature, emission measure, metal abundances, and X-ray bolometric luminosities of our sample and the absorption column density, including interstellar and circumstellar contributions, through X-ray spectroscopy in the following procedure. 
First, we extracted background-subtracted X-ray spectra in the 0.4--10 keV energy range for each of the quiescent and outburst periods.
We refer to the emission during the quiescent periods as the quiescent component.
During the outburst period, we assumed that the observed emission consisted of the quiescent component and an additional outburst component.
Then, we simultaneously fitted the spectra in the quiescent and outburst periods in each observation using the $\chi^2$ method with an absorbed multi-temperature optically thin thermal plasma model. 
The plasma was assumed to be in collisional-ionization equilibrium, as in previous studies \citep[e.g.,][]{2005A&A...429..963O}. 
We adopted the APEC \citep{2001ApJ...556L..91S} and phabs models implemented in the XSPEC package, representing optically thin thermal plasma emission and photoelectric absorption along the line of sight, respectively.
We evaluated the multi-temperature model, expressed as the sum of plasma components at different temperatures for the quiescent and outburst components using the same procedure as \citet{2020ApJ...901...70T}.
We note that the temperatures reported here represent phase-averaged values, obtained by integrating over each identified phase.
We adopted the abundance table presented by \citet{1989GeCoA..53..197A} and assumed that the metal abundances and the absorption column density, $N_H$, were common among the plasma models with different temperatures.
The redshift was fixed to zero in all analyses.
All the parameters unique to the model in the quiescent and outburst component were allowed to vary in the fitting, except when the spectrum of the outburst component had poor statistics, in which case the abundances and absorption column densities were linked between the models for the two components. 
As an exception, the outburst component for V1963~Ori dominates the quiescent component throughout the observed light curve, preventing us from obtaining the quiescent component. We thus did not perform the spectral analysis for this source. 
As another exception, in contrast, although the outburst component for V2108~Ori dominates  the quiescent component throughout the observed light curve, as in V1963~Ori, the quiescent component is not detected because its count rate is consistent with zero within the statistical errors. We therefore fitted the outburst components independently.
We also excluded spectra with insufficient photon statistics either because the outburst duration was too short or because the photon counts were too low to constrain the model parameters.
We adopted a confidence level of 90\% for all the quoted statistical errors. 
In the following sections, when we refer to a temperature for the multi-temperature model, we present a weighted average of the determined multiple temperatures weighted-$kT$ according to the emission measures in the same way as \citet{2022A&A...663A..18B}, while we present the emission measure $EM$ and X-ray luminosities $L_X$ as the sums of all plasma models with different temperatures.

\subsection{Results}
\subsubsection{Timing analysis}

\begin{figure*}[t]
    \includegraphics[width=0.33\textwidth]{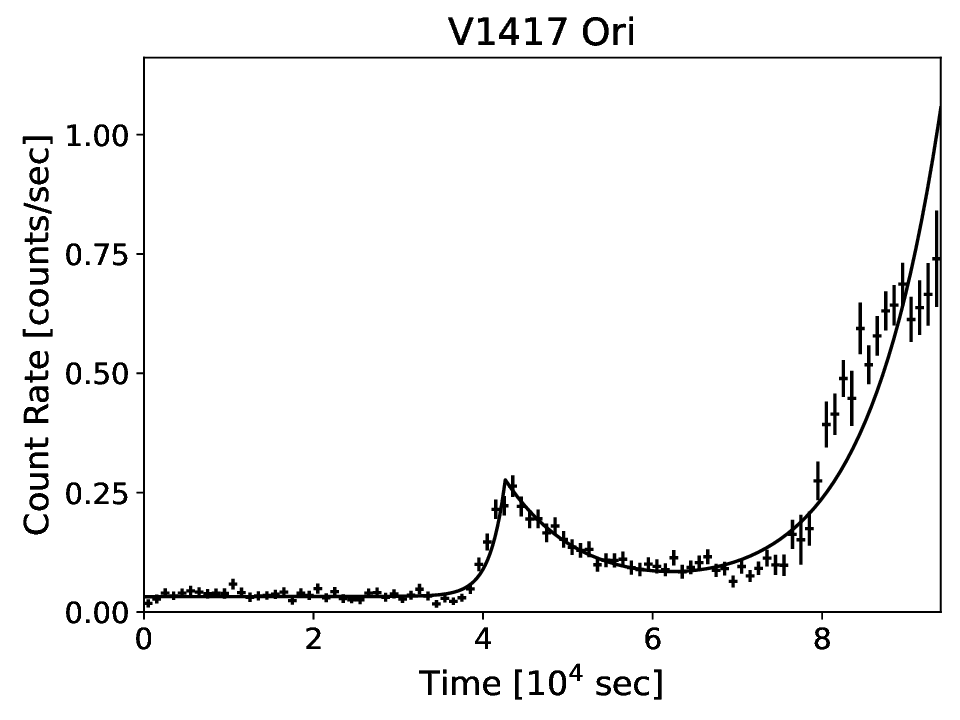} \hfill
    \includegraphics[width=0.33\textwidth]{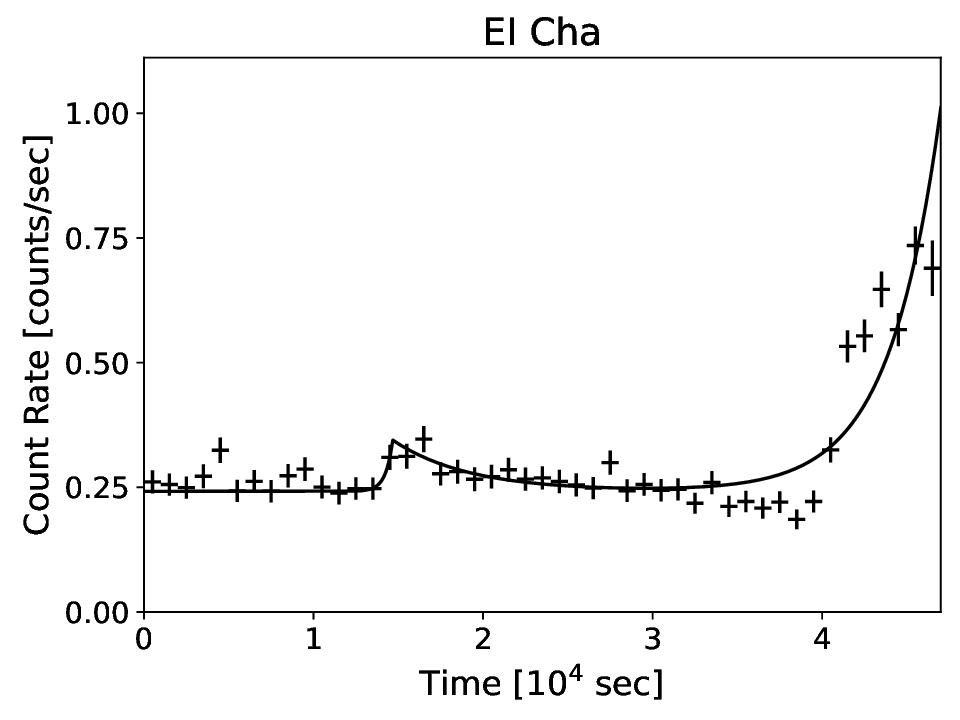} \hfill
    \includegraphics[width=0.33\textwidth]{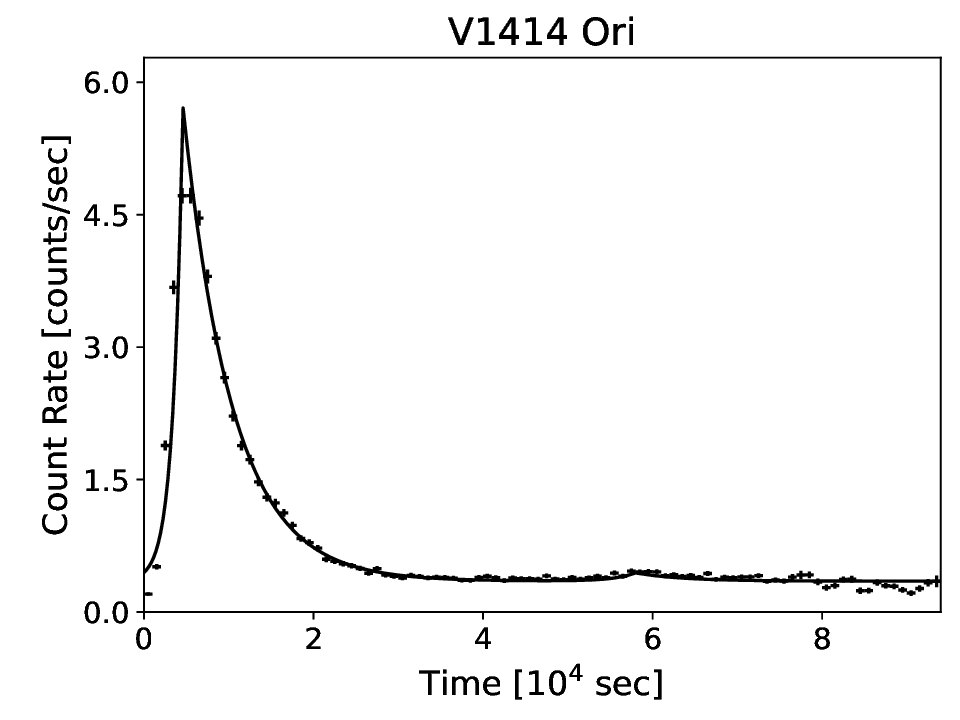} \vspace{3mm}
    \includegraphics[width=0.33\textwidth]{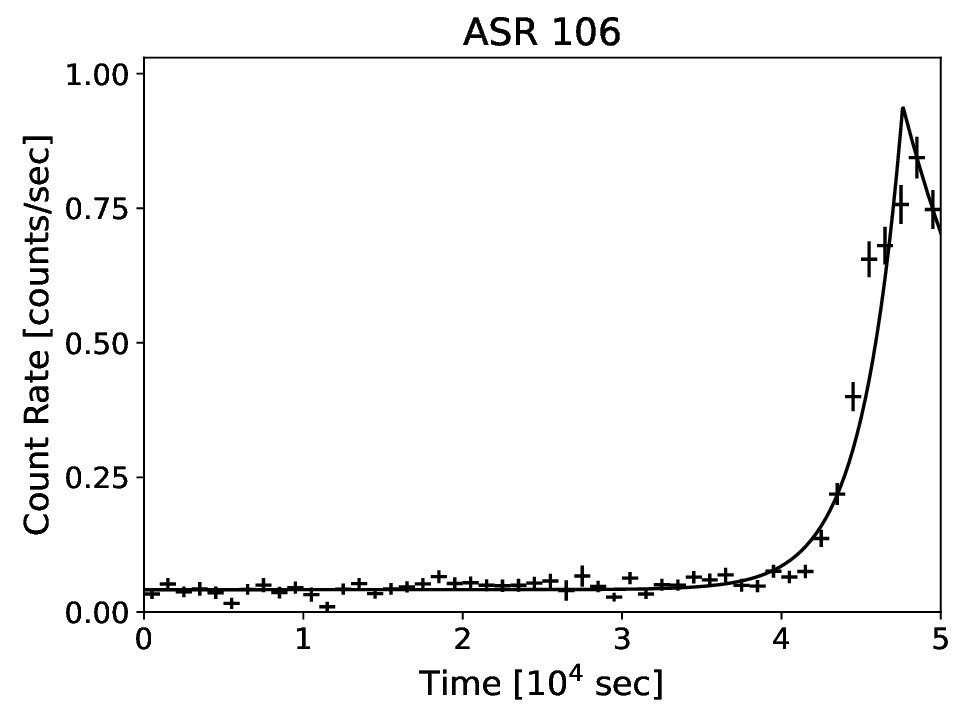} \hfill
    \includegraphics[width=0.33\textwidth]{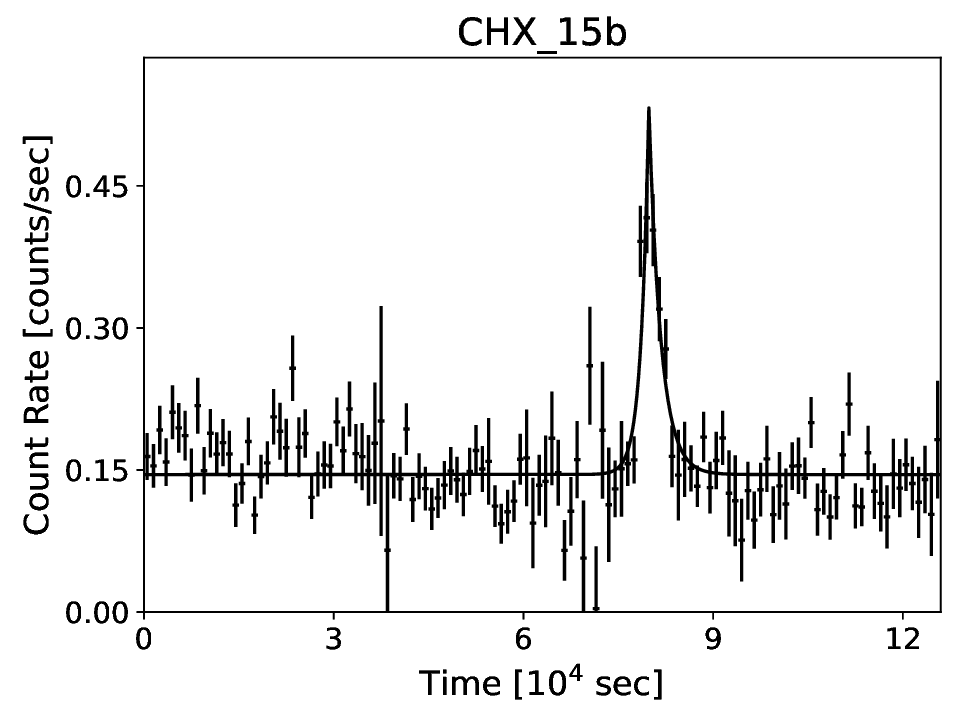} \hfill
    \includegraphics[width=0.33\textwidth]{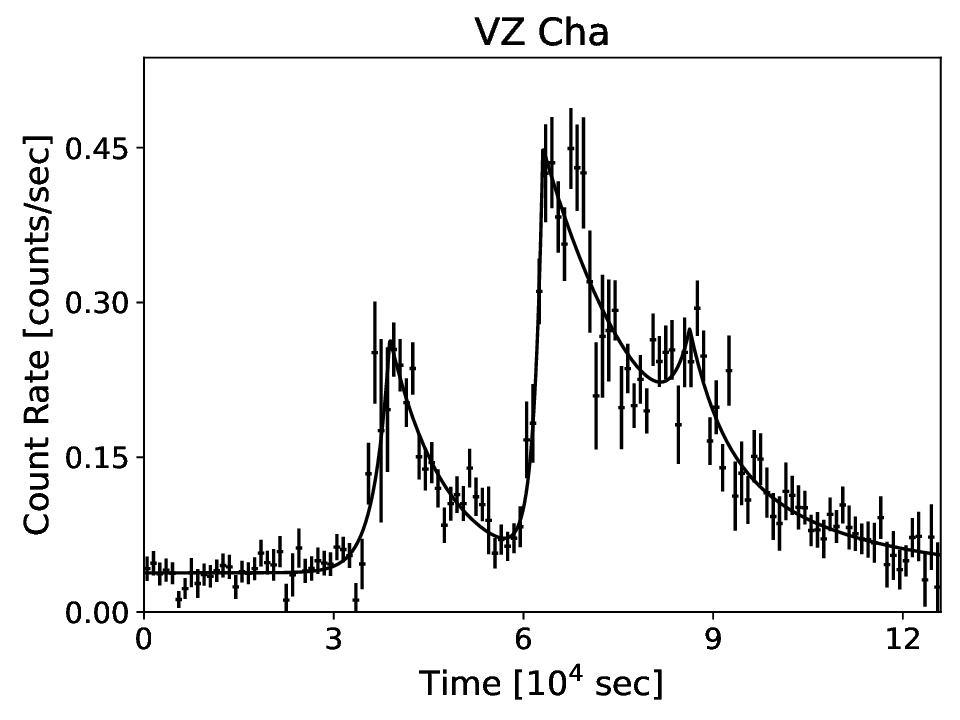}
  \caption{Examples of X-ray light curves of our sample with outbursts during the observations, overlaid with the best-fit light-curve models. The data points represent the observed count rates, and the solid lines indicate the model fits.}
  \label{fig:lc_image}
\end{figure*}

Figures~\ref{fig:lc_image} and \ref{fig:lc_image_a} show the light curves of our sample that exhibited outbursts, and Table~\ref{table:lc} summarizes the fitted parameters.
We detected 50 outbursts in 30 of the 51 YSOs we analyzed. Among the observations, 14 show one outburst, 12 show two outbursts, and 4 show three outbursts within a single exposure, confirming the high variability of X-ray-bright YSOs. Of the detected events, the peak count rates of 31 exceed the quiescent level, indicating large-scale outbursts.
Most light curves are well reproduced by the adopted exponential rise-decay model. In a few cases (e.g., V1417~Ori and EI~Cha), the outbursts cannot be adequately described by this exponential model, and the derived parameters are therefore poorly constrained. In addition, when the rise and/or decay phases are only partially covered or the durations are very short, some parameters remained unconstrained; in these cases, Table~\ref{table:lc} lists the best-fit values where available.
Without the exceptional peak of 5.4~count~s$^{-1}$ in V1414~Ori, the outburst peaks range from 0.05--1~count~s$^{-1}$, while quiescent levels span 0.02--0.9~count~s$^{-1}$. The lower bound of 0.02~count~s$^{-1}$ reflects the selection criterion of our sample. The rise and decay times mostly lie in the ranges $10^{2.5}$--$10^{4.5}$~s and $10^{3}$--$10^{5}$~s, respectively, similar to the distributions reported by \citet{2003PASJ...55..653I}.
The high photon statistics of the observations allowed us to detect rise and decay times shorter than 1000~s, revealing several short-duration outbursts accurately.

For CHX~15b and VZ~Cha we used the same XMM-Newton observations as previous studies \citep{2007A&A...461..669R}. Our timing analysis reveals short-duration outbursts in the former and resolves one quiescent phase and three distinct outbursts in the latter, whereas the data were classified into only two broad activity states previously.
Similarly, for ASR~106 and the members of Mamajek~1, although outbursts were reported in previous analyses of the same datasets \citep{2003A&A...401..543P, 2010A&A...524A..97L}, quantitative parameters were not derived and short-duration events were likely overlooked.
Our timing analysis thus enables the detection of short-duration small-scale outbursts that were difficult to identify in earlier studies. 
By explicitly separating the quiescent and outburst phases, we further enabled the spectroscopic characterization of each phase, as discussed below.

\subsubsection{Spectral analysis}

\begin{figure*}[t]
    \centering
    \includegraphics[width=0.33\textwidth]{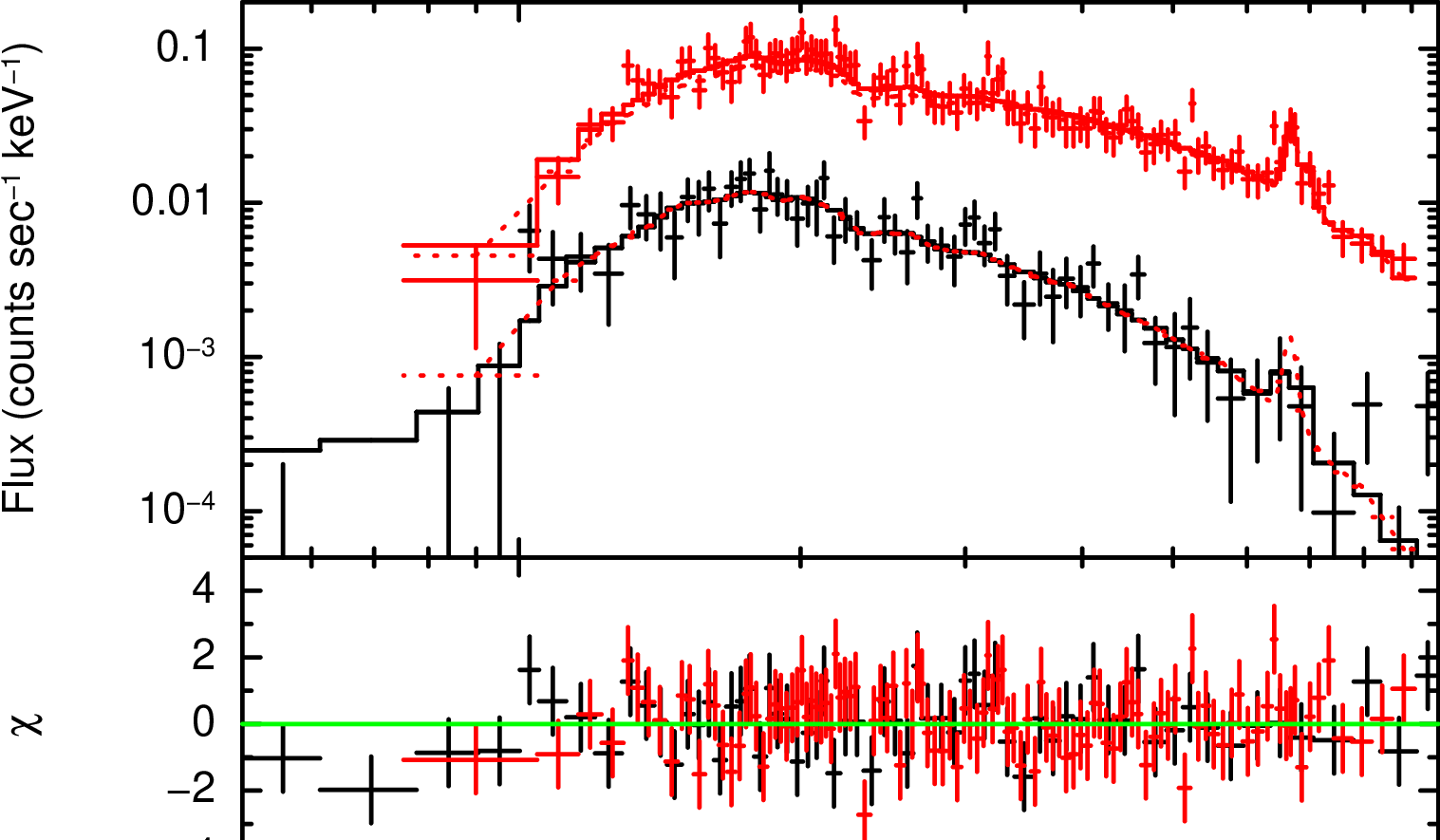}
    \includegraphics[width=0.33\textwidth]{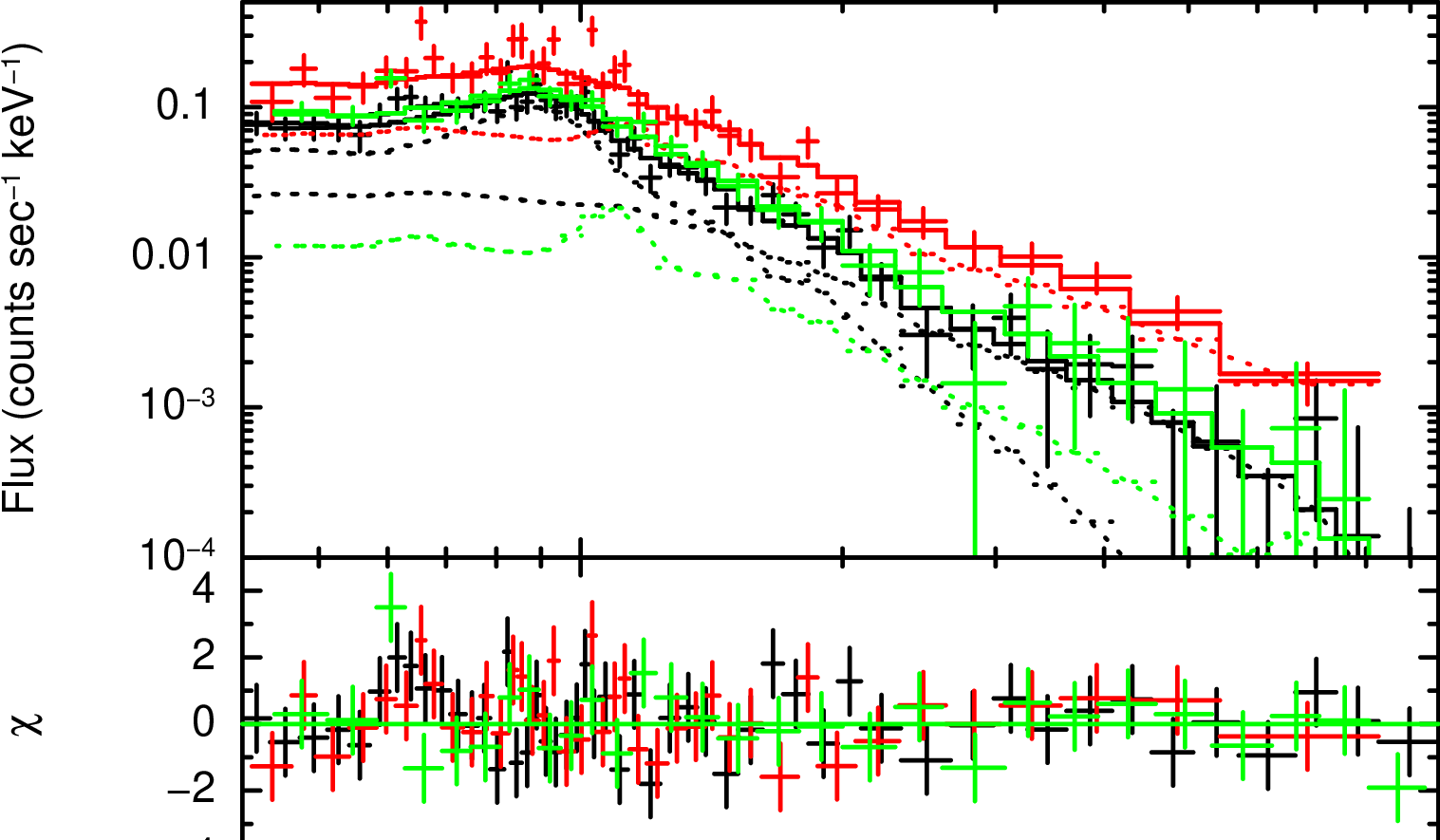}
    \includegraphics[width=0.33\textwidth]{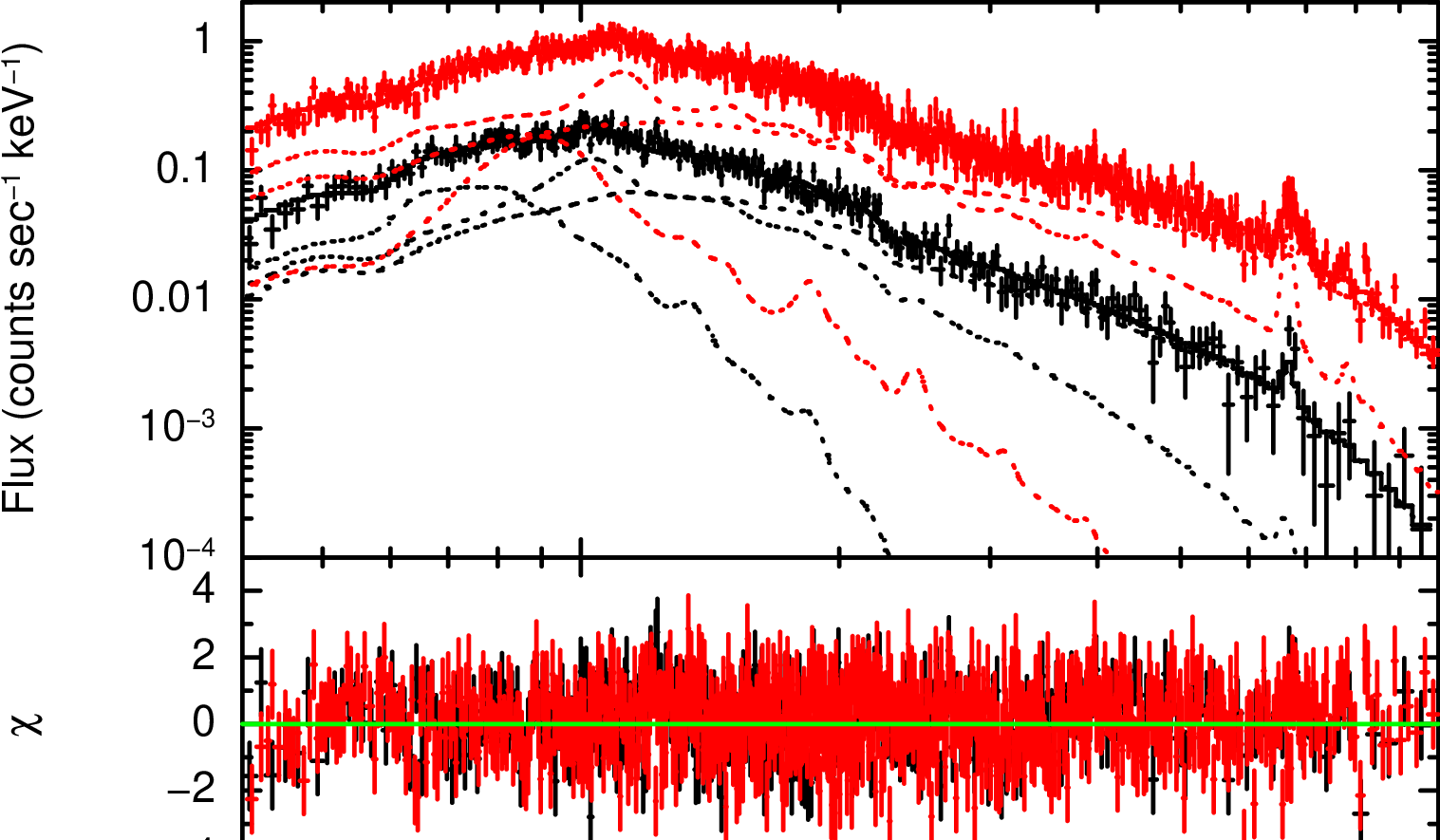} \vspace{3mm}
  \caption{Examples of X-ray energy spectra of ASR~106 (left), Brun~394 (middle), and V1414~Ori (right). The black spectra represent the quiescent phase, and the red and green spectra indicate the outburst phases. In each panel, the upper subpanel shows the flux as a function of energy; the solid lines represent the best-fit multi-temperature plasma models, and the dashed lines indicate the individual thermal plasma models. The lower subpanel shows the $\chi$ values for each data point.}
  \label{fig:sp_image}
\end{figure*}

Figure~\ref{fig:sp_image} shows selected representative spectra overlaid with the best-fit models. 
Table~\ref{table:sp} lists the results of spectral model fitting with multi-temperature optically thin thermal plasma models, summarizing the derived emission-measure-weighted-temperatures, emission measures, X-ray bolometric luminosities, absorption column densities, and metal abundances.
When the photon statistics were insufficient, either preventing reliable parameter constraints through spectral fitting or requiring the parameters of the outburst component to be tied to those of the quiescent component, the corresponding entries are left blank in Table~\ref{table:sp}.
Our spectral analysis shows that quiescent and outburst components can both exhibit multi-temperature plasma structures. 
For the quiescent phase, multi-temperature models are required for the majority of sources (20 two-temperature, 16 three-temperature, and 1 four-temperature cases), whereas a subset of 11 sources is adequately described by a single-temperature model.
In contrast, most outburst components are well reproduced by single-temperature models (39 cases), with only a small fraction requiring additional temperature components (3 two-temperature and 1 three-temperature cases).
For the quiescent components, the weighted temperatures are 0.2--2~keV, with emission measures of $10^{52}$--$10^{55}$~cm$^{-3}$. 
The outburst components reach higher temperatures of 0.5--10~keV, while exhibiting emission measures in a comparable range. 
The X-ray luminosities of the two components span $10^{29}$--$10^{32}$~erg~s$^{-1}$, with absorption column densities of $10^{20}$--$10^{22}$~cm$^{-2}$. 
The metal abundances are typically constrained to 0.1--0.5~$Z_\odot$, and neither absorption nor abundance shows a significant systematic difference between quiescent and outburst components.
Compared with previous spectroscopic studies of YSOs \citep{2003PASJ...55..653I, 2005ApJS..160..319G, 2007A&A...468..485F}, many of the quiescent components in our sample show lower plasma temperatures and higher emission measures.
For the outburst components, the temperature and emission-measure ranges are consistent with these studies. We also detect relatively low-temperature outbursts ($\lesssim 1~{\rm keV}$), however, some of which exhibit high emission measures, such as those observed in V1417~Ori, which have not been clearly characterized in these studies.
These properties are attributable to our high photon statistics and the systematic adoption of multi-temperature spectral models, which allowed us to resolve low-temperature plasma components. 
In addition, the relatively low absorption column densities derived for our sample can further facilitate the detection of such components.
Our results suggest that emission-measure-weighted temperatures and total emission measures in previous studies were biased significantly when low-temperature plasma components were not resolved.
Moreover, the X-ray luminosity of the outburst observed in V1414~Ori reaches $5.4\times10^{32}\ {\rm erg\ s^{-1}}$, which is comparable to that reported in previous studies \citep{1998ApJ...503..894T}, indicating that we have detected one of the most powerful outbursts observed in YSOs.
Our spectroscopic analysis of X-ray-bright YSOs in multiple star clusters suggests that new constraints can be placed on the upper limits of the X-ray luminosities of YSOs in the quiescent and outburst phases \citep[e.g.,][]{2005ApJS..160..319G, 2007A&A...468..425T, 2003PASJ...55..653I}.
These constraints in turn can help us to improve our understanding of the classification and dynamical evolution of disk photoevaporation in protoplanetary disks \citep{2024ApJ...974..281N}.

Despite differences in spectral modeling, the temperatures, luminosities, emission measures, and absorption column densities of CHX 15b and VZ Cha are almost consistent with part of the results reported by \cite{2007A&A...461..669R}. Owing to their fixed metal abundance of $0.5 Z_\odot$, our emission measures are systematically higher.
ASR~106 and EM*~LkHA~270 were analyzed by \citet{2003A&A...401..543P} using the same XMM-Newton data. For ASR~106, the absorption, temperatures, and X-ray luminosities are consistent with our results in the quiescent and outburst phases, whereas EM*~LkHA~270 was not analyzed by separating the quiescent and outburst states, making a direct comparison difficult.
All member stars of Mamajek~1 were analyzed by \citet{2010A&A...524A..97L} using the same XMM-Newton observations. Excluding the newly identified outbursts in this study, our results are largely consistent with a subset of their reported values in terms of plasma temperatures, X-ray luminosities, emission measure ratios, and absorption column densities, despite minor differences in the adopted spectral models.
Although the overall consistency with previous studies is confirmed for the samples discussed above, we adopted the results we derived because of the limited time-resolved analyses and incomplete reporting of physical parameters in previous studies.
HD~287848 remained in a quiescent state throughout the observation, and a two-temperature plasma model was selected. Under these conditions, the adopted procedure is equivalent to that used by \citet{2023RAA....23j5016R}. We therefore adopted the spectral parameters they reported. Although the X-ray luminosity was derived over a different energy band, the values agree within uncertainties, and we likewise adopted the luminosity they reported.

\section{Discussion}\label{sec:discussion}

In this section, we discuss the X-ray emission mechanisms of X-ray-bright YSOs in our sample based on the physical parameters derived from our detailed timing and spectral analyses.
We first examine the magnetic reconnection and magnetospheric accretion scenarios from multiple perspectives using the plasma properties obtained in this study.
We then introduce an additional diagnostic by investigating correlations between the quiescent and outburst phases.
In the following discussion, we restrict our comparison to the subset of our sample for which the physical parameters are well constrained, with determined upper and lower bounds.

\subsection{Comparison with scaling relations for magnetic reconnection and magnetospheric accretion}

In this section, we compare the observed X-ray properties with theoretical scaling relations for magnetic reconnection and magnetospheric accretion. 
For the magnetic reconnection scenario, we compared the observed quantities with the scaling relations discussed by \citet{1999ApJ...526L..49S,2002ApJ...577..422S} and \citet{2003PASJ...55..653I}. 
The observed temperatures, emission measures, and timescales in the quiescent and outburst phases are broadly consistent in an order-of-magnitude sense with the ranges expected from magnetic reconnection scaling relations. 
In particular, the inferred loop lengths are comparable to the large magnetic loops reported for YSO flares by \citet{2003PASJ...55..653I}. 
The similar magnetic loop length scales inferred for the quiescent phase can indicate that large-scale magnetic structures are present not only during outbursts, but also in the underlying quiescent coronae.
For the magnetospheric accretion scenario, we compared the observed X-ray luminosities and plasma temperatures with the scaling relations of \citet{1998ApJ...509..802C}. 
This comparison was only possible for the 19 sources for which stellar masses and radii are available from Gaia DR3. 
For this subset, the mass-to-radius ratios, $M_\ast/R_\ast$, typically lie in the range of 0.4--1.0 in solar units. 
For typical mass-accretion rates of YSOs, the observed X-ray luminosities are broadly comparable to those expected from magnetospheric accretion in an order-of-magnitude sense.
In contrast, the expected accretion-shock temperatures are $\lesssim 0.3$ keV, and for most of these sources, even the lowest-temperature component in the multi-temperature spectral model is hotter than the shock temperature estimated from the stellar masses and radii.
This result suggests that for most objects in this subsample, the magnetospheric accretion scenario alone cannot explain the X-ray-emitting plasma detected in the XMM-Newton/EPIC-pn energy band.
We note, however, that this discussion is limited to the objects for which model-dependent stellar masses and radii can be estimated from Gaia DR3.
Motivated by the expected temperature range of magnetospheric accretion shocks, we reconstructed the $EM$--$kT$ relation after excluding model components with $kT \lesssim 0.3$ keV. 
With this separation, the quiescent components show a clearer positive correlation in the $EM$--$kT$ distribution, with a Spearman correlation coefficient of 0.51.
The inferred magnetic loop lengths are then distributed around $10^{12}$ cm, corresponding to $\sim 0.07$ AU. 
If the $kT \lesssim 0.3$ keV model components include contributions from magnetospheric accretion shocks, the remaining emission might trace large-scale magnetic loops extending to the scale of the inner protoplanetary disk \citep{2008ApJ...688..437G}. 
This suggests that X-ray-bright YSOs can maintain large-scale magnetic structures, while accretion shocks and magnetic reconnection can both contribute to their X-ray emission.

\subsection{Neupert effect and time evolution of $kT$ and $EM$} 

\begin{figure*}[t]
    \includegraphics[width=0.33\textwidth]{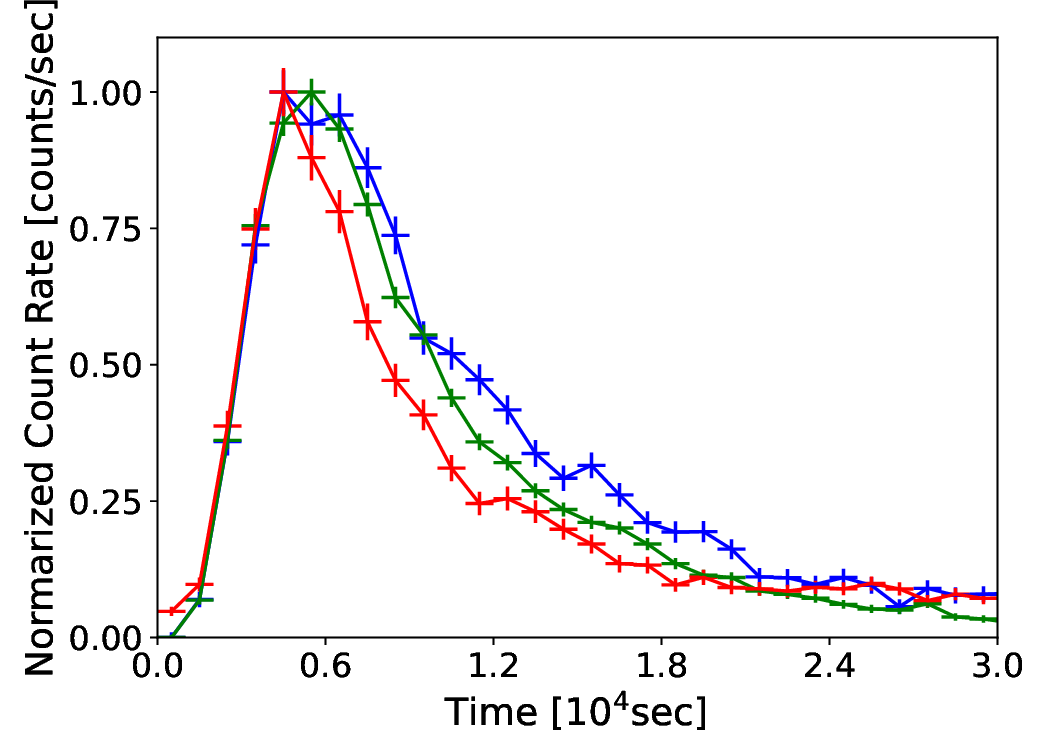} \hfill
    \includegraphics[width=0.33\textwidth]{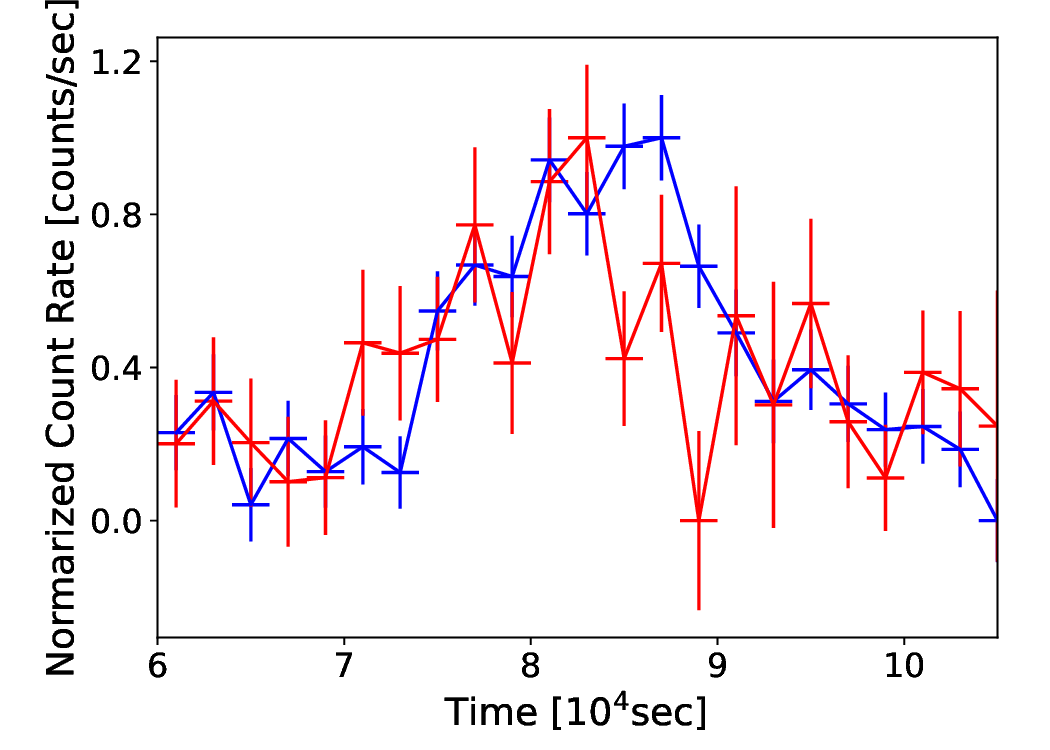} \hfill
    \includegraphics[width=0.33\textwidth]{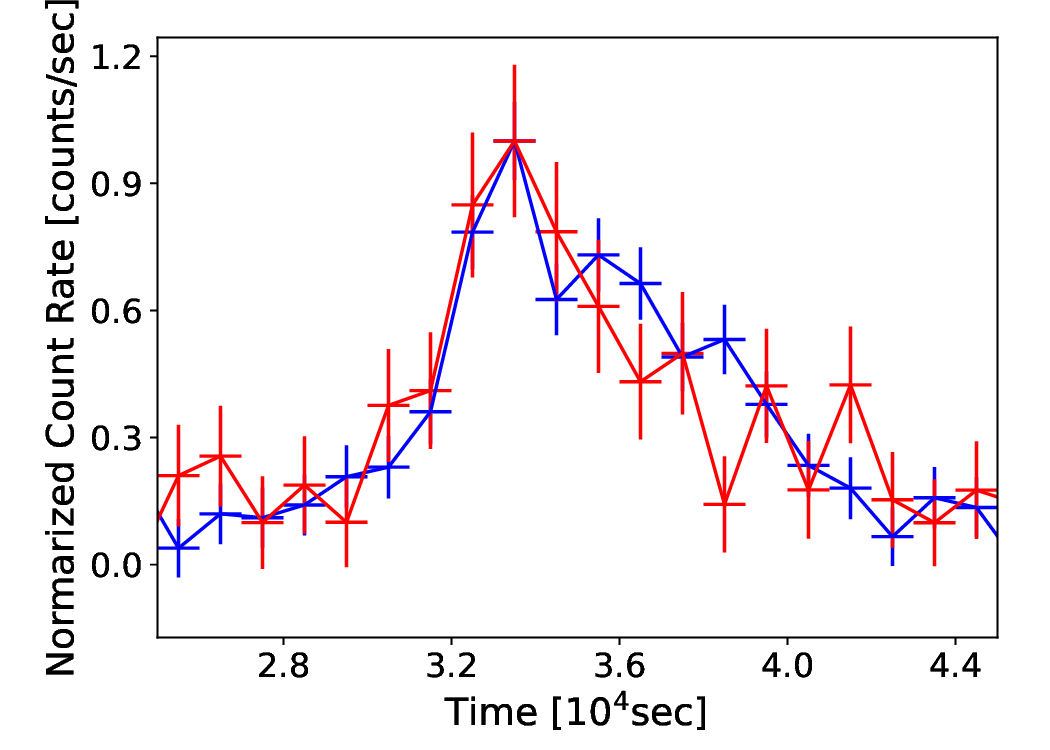} \vspace{3mm}
    \includegraphics[width=0.33\textwidth]{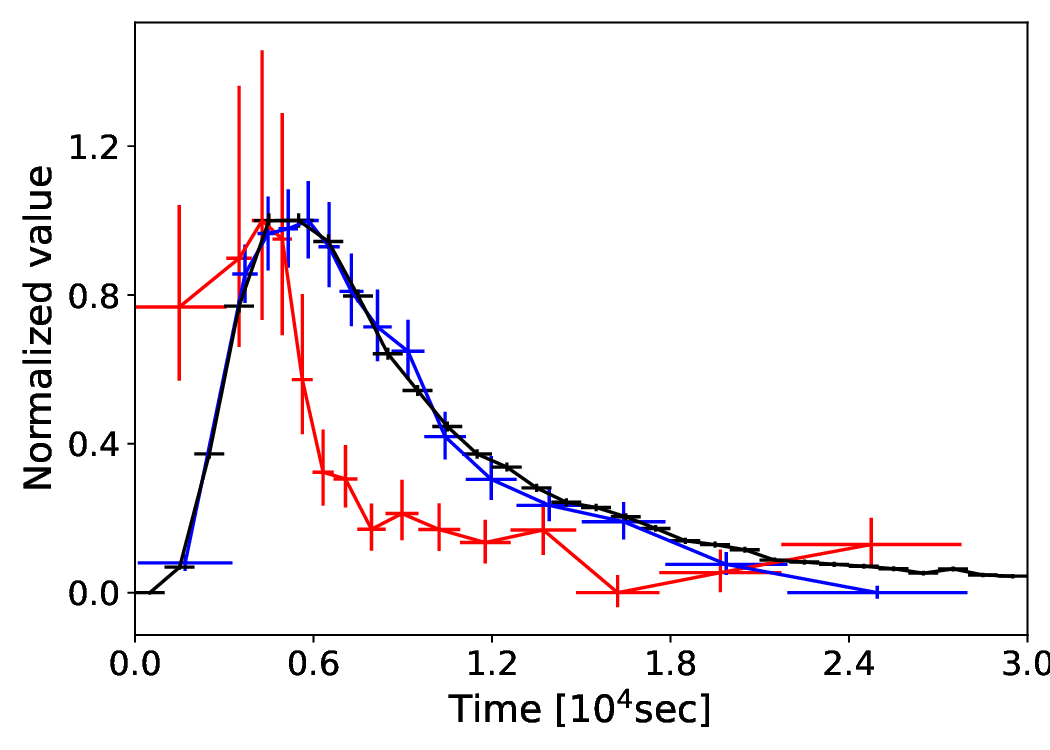} \hfill
    \includegraphics[width=0.33\textwidth]{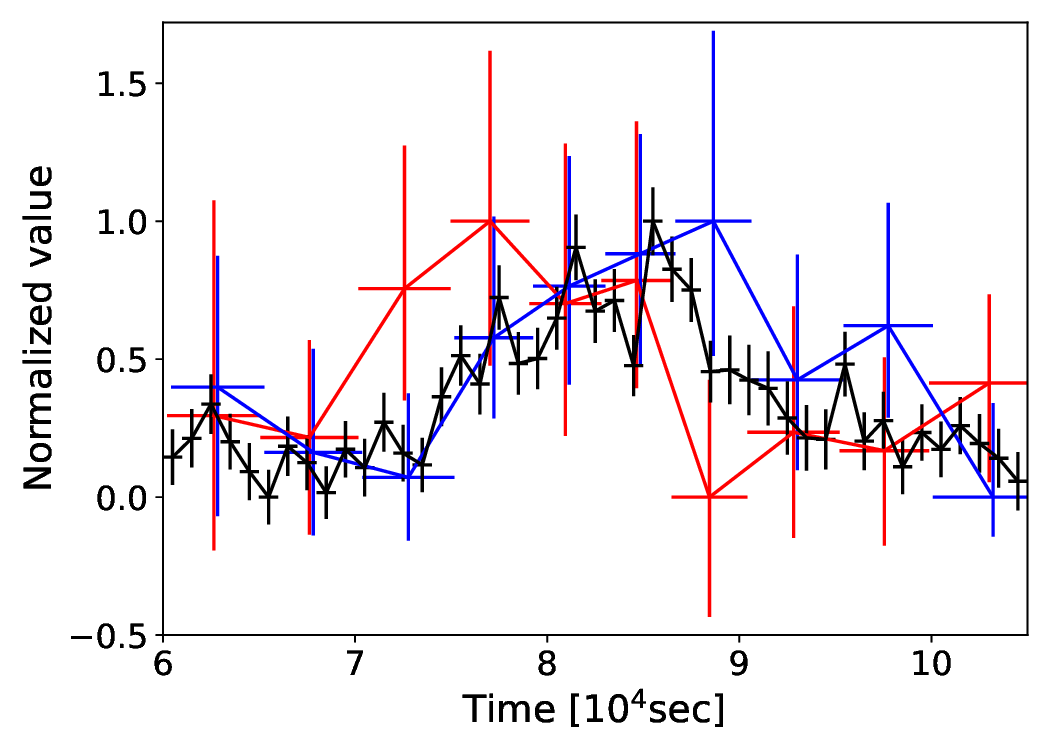} \hfill
    \includegraphics[width=0.33\textwidth]{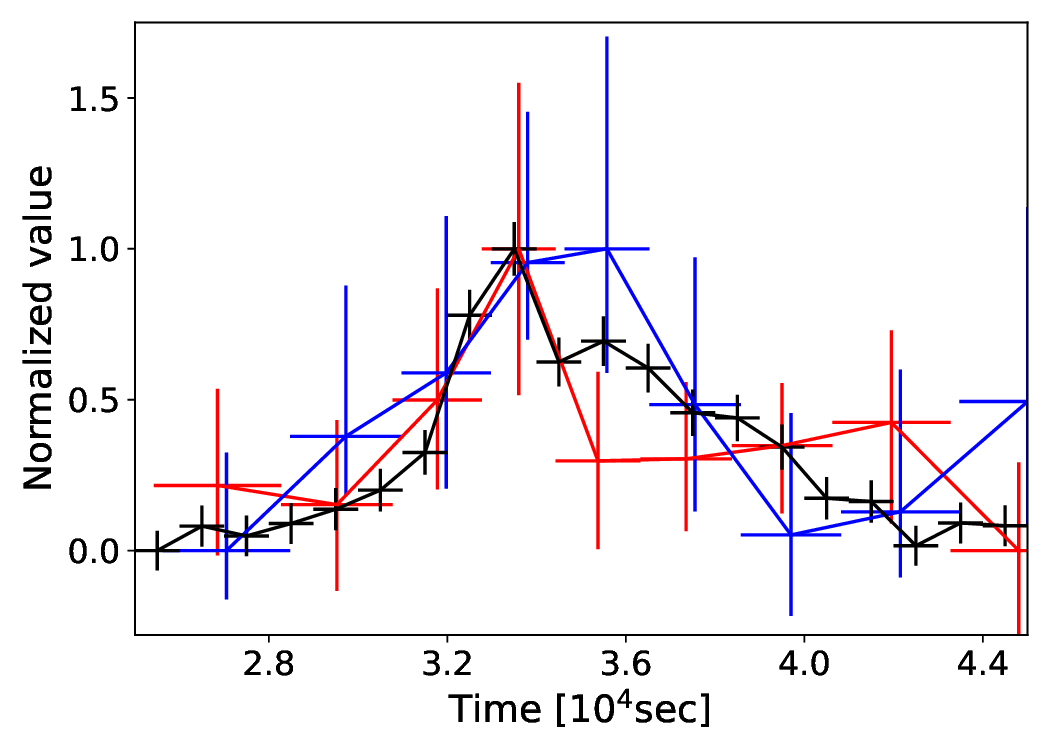} 
  \caption{Example light curves and time evolution of the emission measure and temperature for the three sources with the highest photon statistics of those that exhibited outbursts during the observations, V1414~Ori (left panels), V1878~Ori (middle panels), and EP~Cha (right panels).
  The upper panels show light curves in the 0.4--1 keV (blue), 1--3 keV (green), and 3--5 keV (red) bands for V1414~Ori, and in the 0.4--2 keV (blue) and 2--5 keV (red) bands for V1878~Ori and EP~Cha. 
  The lower panels show the light curve in the total energy band (0.4--5 keV; black), together with the time evolution of the temperature (red) and emission measure (blue).}
  \label{fig:time_vari}
\end{figure*}

To investigate observational signatures of magnetic reconnection-driven outbursts, specifically the Neupert effect and the temporal evolution of $kT$ and $EM$, we focused on the three sources with the highest photon statistics of those that exhibited outbursts during the observations: V1414~Ori, V1878~Ori, and EP~Cha. 
To investigate the Neupert effect, we extracted light curves in three energy bands of 0.4--1~keV, 1--3~keV, and 3--5~keV for V1414~Ori, and for EP~Cha and V1878~Ori, we used two energy bands of 0.4--2~keV and 2--5~keV.
We binned each light curve to ensure sufficient photon statistics. For V1414~Ori and EP~Cha, we applied a binning with 1,000 seconds per bin to all the light curves, and for V1878~Ori, we applied 2,000 seconds.
Next, we constructed the temporal evolution of the $kT$ and $EM$ as follows.
First, we defined the time bins, requiring that each bin contained 2,000 counts at least for the V1414~Ori data and 1,000 counts at least for the V1878~Ori and EP~Cha data.
Then, we extracted an energy spectrum for each time bin, fit it with an absorbed single-temperature optically thin thermal plasma model, and obtained the $kT$ and $EM$. 
In the spectral fitting, we allowed the absorption and metal abundance to vary while fixing the redshift to zero.
Figure \ref{fig:time_vari} presents the derived light curves in the respective energy bands and the temporal evolution of $kT$ and $EM$. 
For V1414~Ori, the light curves provide one of the few clear observational signatures of the Neupert effect in YSOs. 
For V1878~Ori, the light-curve behavior suggests the Neupert effect.
In EP~Cha, a similar tendency is present but appears to be weaker, which can be attributed to the lower count rate of the outburst component relative to the quiescent emission.
Previous studies have reported Neupert-like behavior through comparisons with ultraviolet emission \citep{2007A&A...468..379A} and the Neupert effect in the hard X-ray band \citep{2019ApJ...882...72V}. In contrast, our results clearly demonstrate the Neupert effect in low-temperature plasma, including energies below 1~keV, providing a rare example of this behavior in YSOs.
In the temporal evolution of $kT$ and $EM$ for V1414~Ori, we confirm that the rise and subsequent decline of $kT$ precedes the variation in $EM$. Although this behavior has been reported previously \citep{2005ApJS..160..469F, 2007A&A...468..485F}, the decay-phase offset is more significantly resolved in this unusually bright outburst.
For V1878~Ori, a similar behavior is observed, with an increase in $kT$ prior to the peak of $EM$.
In EP~Cha, although the result is not statistically significant, the peak of $EM$ is delayed relative to that of $kT$.
These temporal features provide clear observational evidence that the particularly X-ray-bright YSO outbursts in our sample exhibit a time evolution that is consistent with magnetic reconnection-driven processes.

\subsection{X-ray activity}\label{sec:rossby}

Late-type stars, including the Sun, are known to exhibit magnetic reconnection-driven outbursts, which are thought to arise from efficient magnetic field generation and amplification by stellar dynamos \citep[e.g.,][]{1955ApJ...122..293P, 2004A&ARv..12...71G, 2014ARA&A..52..251C}. 
Motivated by this, we investigated whether our sample followed the relation between the Rossby number and X-ray activity established for late-type stars \citep{2011ApJ...743...48W}.
The Rossby number, widely used as an indicator of magnetic activity, is defined as the ratio of the stellar rotation period to the convective turnover time.
We obtained the stellar rotation periods of our sample from observations by the optical space telescope TESS \citep{2019ASPC..523..453S}, adopting the average values of the available measurements.
The uncertainties were taken as the standard deviations of the measured periods.
We estimated the convective turnover times by comparing the effective temperatures and luminosities listed in Gaia DR3 with the evolution tracks presented by \citet{2007JASS...24....1J}, following the same procedure as in \citet{2024A&A...691A.304Y}. 
Table~\ref{table:rossby} lists the derived Rossby numbers and the ratios of the X-ray to bolometric luminosities, $L_{\rm X}/L_{\rm bol}$  for the quiescent and outburst components, an indicator of X-ray activity.
The Rossby numbers of all YSOs in our sample except for the high-mass star HD~36959 correspond to the saturated regime ($< 0.1$).
In late-type stars in the saturated regime, $L_{\rm X}/L_{\rm bol}$ typically lies in the range of $10^{-4}$--$10^{-2}$ \citep{2011ApJ...743...48W}. The quiescent and outburst components of our sample fall within this range.
We estimated $L_{\rm X}/L_{\rm bol}$ for the YSOs in our sample for which Rossby numbers could not be derived owing to insufficient information or limited data statistics. 
The average values of $L_{\rm X}/L_{\rm bol}$ for the quiescent and the outburst components are $10^{-3.0}$, and $10^{-3.3}$, respectively, with dispersions of $10^{0.8}$ and $10^{0.4}$.
Even though Rossby numbers are not available for most of our sample, the $L_{\rm X}/L_{\rm bol}$ values for most of our sample are consistent with the saturation regime in the quiescent and outburst states.
The concentration of sources at high X-ray activity levels is reasonable because our sample consists of X-ray-bright YSOs.
This tendency suggests that the X-ray activity is in the saturated regime, which is thought to correspond to a state in which the stellar dynamo is saturated or the stellar surface is nearly fully covered by magnetically active regions \citep{2011MNRAS.411.2099J}.
In conclusion, our results indicate a similarity between X-ray-bright YSOs and late-type stars in their X-ray activity, with the former likely residing in the saturated regime. 
This implies efficient magnetic loop generation and suggests that, as in late-type stars, their X-ray emission is closely linked to magnetic reconnection.
This similarity is consistent with the photometric study of \citet{2016A&A...589A.113A}, who derived X-ray luminosities averaged over the whole exposure time and showed that low-mass YSOs follow a relation similar to that of late-type stars.
In contrast, by explicitly isolating the quiescent emission and performing spectral analysis, we provide a spectroscopic confirmation based on more accurately determined X-ray luminosities and demonstrate that quiescent and outburst phases independently reside in the saturated regime.

\begin{table}[ht!]
    \caption{Rossby number and X-ray activity in the quiescent and outburst components.}
    \label{table:rossby}
    \centering
    \begin{tabular}{llcc}
        \hline\hline
        Star & phase & Rossby number & $L_{\rm X}/L_{\rm bol}$ [$10^{-3}$] \\ \hline
        Hen~3-545 & quiescent & 0.024 & $0.40_{-0.08}^{+0.13}$ \\
         & outburst &   & $0.060_{-0.016}^{+0.017}$ \\
        V1740~Ori & quiescent & 0.044 & $1.5_{-0.3}^{+0.6}$ \\
        HD~36959 & quiescent & 0.23 & $0.00060_{-0.00011}^{+0.00016}$ \\
        EI~Cha & quiescent & $0.025 \pm 0.002$~\tablefootmark{a} & $0.95_{-0.11}^{+0.17}$ \\
         & outburst~1 &   & $0.12_{-0.07}^{+0.10}$ \\
         & outburst~2 &   & $1.5_{-0.3}^{+0.3}$ \\
        EL~Cha & quiescent & $0.0075 \pm 0.0012$~\tablefootmark{a} & $0.59_{-0.07}^{+0.06}$ \\
         & outburst~1 &   & $0.45_{-0.19}^{+0.51}$ \\
         & outburst~2 &   & $0.74_{-0.30}^{+0.30}$ \\
        EO~Cha & quiescent & $0.023 \pm 0.006$~\tablefootmark{a} & $0.50_{-0.04}^{+0.10}$ \\
        EP~Cha & quiescent & $0.039 \pm 0.016$~\tablefootmark{a} & $2.9_{-1.2}^{+1.2}$ \\
         & outburst~1 &   & $0.26_{-0.06}^{+0.08}$ \\
         & outburst~2 &   & $0.28_{-0.07}^{+0.06}$ \\
         & outburst~3 &   & $0.48_{-0.06}^{+0.06}$ \\
        \hline
    \end{tabular}
    \raggedright
    \tablefoottext{a}{Because of multiple measurements from TESS, the uncertainty is given as the standard deviation of the measurements.}
\end{table}

\subsection{Correlation between physical properties in the quiescent and outburst phases}\label{sec:corr}

\begin{figure}[htbp]
  \begin{center}
  \includegraphics[width=\linewidth]{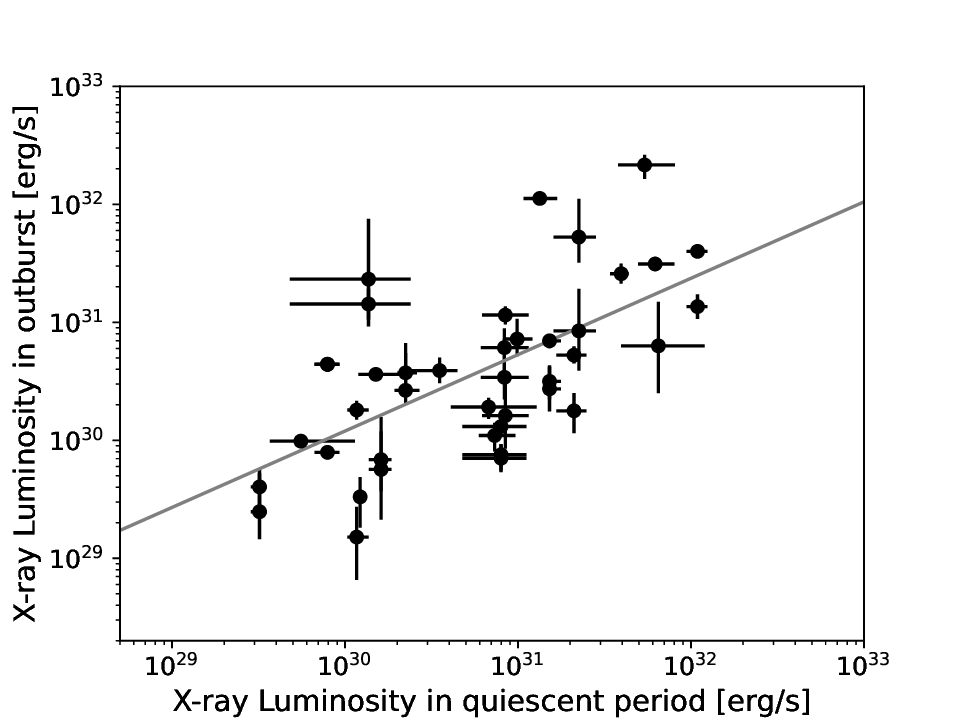}
  \end{center}
  \caption{Correlation between the X-ray luminosities of the quiescent and outburst components. Only events with constrained upper and lower bounds are shown. The gray line indicates the best-fit power-law approximation, given by $y = bx^a$, where $a = 0.65$ and $b = 10^{11}$.} 
  \label{fig:q-o_lumi}
\end{figure}

By explicitly separating the quiescent and outburst phases and deriving their physical parameters in detail, we were able to examine the correlations of the physical quantities of these two components from a new observational perspective.
The temperature is weakly positively correlated, with a correlation coefficient of 0.24.
In contrast, the emission measure and the X-ray luminosity are strongly positively correlated, with correlation coefficients of 0.51 and 0.64, respectively.
Figure~\ref{fig:q-o_lumi} shows the correlation between the X-ray luminosities for the quiescent and outburst components.
Data points corresponding to high X-ray luminosities in the quiescent component and low X-ray luminosities in the outburst component are less likely to be plotted because outbursts are difficult to detect in these cases.
However, the absence of sources with low X-ray luminosities in the quiescent component and high X-ray luminosities in the outburst component suggests that this parameter space is not merely undersampled, but might instead be regulated by an uncertain physical mechanism that suppresses such combinations.
A similar behavior has been reported for the Sun, where the flare magnitude correlates with the level of the non-flaring soft X-ray background \citep{2015SpWea..13..286W}.
Therefore, the observed correlations of the quiescent and outburst components, together with their similarity to solar behavior, suggest that magnetic reconnection-driven processes likely drive the X-ray emission in the two phases of our sample.

\section{Conclusions}\label{sec:conclusions}

We performed a systematic analysis of XMM-Newton X-ray data for a sample of 51 X-ray-bright YSOs to derive their X-ray properties and to investigate their emission mechanisms through an approach from multiple perspectives based on timing and spectroscopic analyses, in terms of magnetic reconnection and magnetospheric accretion scenarios.
Through a timing analysis of observations with high photon statistics, we successfully detected several short-duration outbursts, quantitatively identified and separated the quiescent and outburst phases, and performed X-ray spectroscopy for each phase using multi-temperature plasma models.

Our timing analysis identified 50 outbursts in 30 YSOs in our sample. The outburst peak count rates typically range from 0.05 to 1~count~s$^{-1}$, while quiescent count rates span 0.02--0.9~count~s$^{-1}$. The rise and decay times mostly fall in the ranges $10^{2.5}$--$10^{4.5}$ and $10^{3}$--$10^{5}$~s, respectively. Multiple outbursts were detected within single observations in approximately half of the sample, demonstrating that X-ray-bright YSOs are highly variable and magnetically active systems.
Our multi-temperature spectral analysis revealed that quiescent and outburst components commonly exhibit multi-temperature plasma structures. The quiescent components are characterized by emission-measure-weighted temperatures of 0.2--2~keV, whereas the outburst components reach higher temperatures of 0.5--10~keV. In contrast, the emission measures ($10^{52}$--$10^{55}$~cm$^{-3}$) and X-ray luminosities ($10^{29}$--$10^{32}$~erg~s$^{-1}$) span comparable ranges in both phases.
Low-temperature plasma components of $\lesssim$1~keV were detected in the quiescent and outburst phases, some of which exhibit relatively high emission measures.
The detection suggests that the emission-measure-weighted temperature and the total emission measure are significantly biased when these components are not properly accounted for.

We compared the relations among the parameters estimated from the timing and spectral analyses with theoretical models proposed in previous studies. 
The observed properties are broadly consistent with the magnetic reconnection scenario. 
In contrast, the comparison based on the Gaia DR3 stellar parameters suggests that magnetospheric accretion alone cannot explain most of the fitted X-ray temperatures, although the coolest model components with $kT \lesssim 0.3$ keV might include an accretion-shock contribution. 
After we separated these low-temperature components, the remaining quiescent components were more clearly positively correlated in the $EM$--$kT$ plane, with inferred magnetic loop lengths distributed around $10^{12}$ cm, or $\sim 0.07$ AU. 
These results suggest that magnetic reconnection is the primary origin of the X-ray emission in X-ray-bright YSOs, while magnetospheric accretion can contribute to the coolest components.
Observational characteristics, including the detection of a Neupert effect, the temporal evolution of temperature and emission measure, and the relation of the Rossby number and X-ray activity, also support the magnetic reconnection scenario.
We introduced an additional diagnostic by examining correlations between the physical parameters of the quiescent and outburst phases, enabled by the explicit phase separation achieved in this study. We found positive correlations in temperature, emission measure, and X-ray luminosity between the two components. These correlations, together with their similarity to solar behavior, suggest that magnetic reconnection-driven processes likely govern the X-ray emission in the quiescent and outburst phases of our sample.
Our results are important for the ionization and chemical evolution of inner protoplanetary disks \citep{2024ApJ...976...25W, 2023ApJ...956..103W}, and the detailed characterization of outbursts can provide observational upper limits on the role of outburst heating in the production of CAIs in meteorites in the early Solar System \citep{2021SciA....7.8329B}.

\begin{acknowledgements}
This work was financially supported by JST SPRING, Grant Number JPMJSP2125. K.S. would like to take this opportunity to thank the “THERS Make New Standards Program for the Next Generation Researchers.” 
This work was supported by JST FOREST Program, Grant Number JPMJFR2369 (IM) and Grants-in-Aid for Scientific Research (KAKENHI) of the Japanese Society for the Promotion of Science (JSPS, grant Nos. JP25K01012 and JP22K18274 (IM), and JP21H04492 and JP25K01041 (ST)).
Based on observations obtained with XMM-Newton, an ESA science mission with instruments and contributions directly funded by ESA Member States and NASA. 
This research has made use of the WEBDA database operated at the Institute of Astronomy of the University of Vienna, Austria. 
This research has made use of the SIMBAD database, operated at CDS, Strasbourg, France. 
This work has made use of data from the European Space Agency (ESA) mission Gaia (\texttt{https://www.cosmos.esa.int/gaia}), processed by the Gaia Data Processing and Analysis Consortium (DPAC, \texttt{https://www.cosmos.esa.int/web/gaia/dpac/consortium}). Funding for the DPAC has been provided by national institutions, in particular the institutions participating in the Gaia Multilateral Agreement. 
This paper includes data collected by the TESS mission, which are publicly available from the Mikulski Archive for Space Telescopes (MAST).
\end{acknowledgements}

\bibliographystyle{aa}
\bibliography{citation}

\begin{appendix}

\onecolumn
\section{Sample information}
This appendix provides the basic information of our sample obtained from the 4XMM-DR14 and Gaia DR3 catalogs.

\begin{longtable}{lllcccc}
\caption{Our sample selected from the 4XMM-DR14 catalog}\\
\label{table:analyzed_star}\\
\hline\hline
Star cluster & Star & 4XMM-DR14 ID & R.A.$_{\rm J2000}$ & Decl.$_{\rm J2000}$ & Obs. ID & Counts \\
  &   &   & [deg] & [deg] &   & \\
\hline
\endfirsthead
\caption{continued.}\\
\hline\hline
Star cluster & Star & 4XMM-DR14 ID & R.A.$_{\rm J2000}$ & Decl.$_{\rm J2000}$ & Obs. ID & Counts \\
  &   &   & [deg] & [deg] &   & \\
\hline
\endhead
\hline
\endfoot
Chamaeleon~I & CHX 15b & 4XMM~J110940.1-762839 & 167.417 & -76.477 & 0300270201 & 4216 \\
  & Hen 3-545 & 4XMM~J105906.9-770140 & 164.779 & -77.027 & 0067140201 & 9282 \\
  & VZ Cha & 4XMM~J110923.6-762320 & 167.349 & -76.389 &  0300270201 & 4216 \\
NGC 1976 & V557 Ori & 4XMM~J053513.3-050920 & 83.806& -5.156 & 0093000101 & 2892 \\
  & V569 Ori & 4XMM~J053543.2-050917 & 83.930 & -5.155 & 0134531601 & 5620 \\
  & V575 Ori & 4XMM~J053551.1-050709 & 83.963 & -5.119 & 0134531601 & 2967 \\
  & V984 Ori & 4XMM~J053531.2-055834 & 83.880 & -5.976 & 0690200501 & 4193 \\
  & V1044 Ori & 4XMM~J053416.4-053645 & 83.568 & -5.613 & 0403200101 & 4545 \\
  & V1221 Ori & 4XMM~J053428.2-061024 & 83.618 & -6.174 & 0503560701 & 3967 \\
  & V1414 Ori & 4XMM~J053357.7-054006 & 83.491 & -5.668 & 0605900101 & 43222 \\
  & V1417 Ori & 4XMM~J053401.7-054651 & 83.507 & -5.781 & 0605900101 & 4281 \\
  & V1553 Ori & 4XMM~J053530.2-055116 & 83.876 & -5.855 & 0690200501 & 4586 \\
  & V1667 Ori & 4XMM~J053346.1-053427 & 83.442 & -5.574 & 0605900101 & 2733 \\
  & V1740 Ori & 4XMM~J053529.0-050604 & 83.871 & -5.101 & 0093000101 & 2838 \\
  & V1760 Ori & 4XMM~J053551.6-050809 & 83.965 & -5.136 & 0093000101 & 6417 \\
  & V1878 Ori & 4XMM~J053042.6-043501 & 82.678 & -4.584 & 0763720401 & 23803 \\
  & V1963 Ori & 4XMM~J053427.7-053155 & 83.615 & -5.532 & 0403200101 & 18871 \\
  & V2087 Ori & 4XMM~J053457.6-061831 & 83.740 & -6.309 & 0503560701 & 17996 \\
  & V2098 Ori & 4XMM~J053459.5-062139 & 83.748 & -6.361 & 0503560701 & 3567 \\
  & V2108 Ori & 4XMM~J053501.3-054113 & 83.755 & -5.687 & 0403200101 & 2718 \\
  & AR Ori & 4XMM~J053554.0-050415 & 83.975 & -5.071 & 0093000101 & 2557 \\
  & PR Ori & 4XMM~J053624.9-061732 & 84.104 & -6.292 & 0089940301 & 11032 \\
  & WW Ori & 4XMM~J053403.0-053656 & 83.513 & -5.616 & 0605900101 & 5243 \\
  & YZ Ori & 4XMM~J053453.0-050327 & 83.721 & -5.058 & 0093000101 & 2721 \\
  & Brun 213 & 4XMM~J053414.4-052816 & 83.560 & -5.471 & 0403200101 & 6009 \\
  & Brun 302 & 4XMM~J053433.9-052824 & 83.641 & -5.473 & 0403200101 & 6845 \\
  & Brun 311 & 4XMM~J053435.1-053210 & 83.646 & -5.536 & 0403200101 & 7967 \\
  & Brun 394 & 4XMM~J053445.9-055851 & 83.691 & -5.981 & 0690200501 & 4469 \\
  & Brun 656  & 4XMM~J053521.3-051212 & 83.839 & -5.204 & 0093000101 & 19347 \\
  & Brun 684  & 4XMM~J053521.9-055107 & 83.841 & -5.852 & 0112660101 & 4876 \\
  & BD-06 1236 & 4XMM~J053509.0-061420 & 83.788 & -6.239 & 0503560701 & 2727 \\
  & HD36959 & 4XMM~J053501.0-060033 & 83.754 & -6.009 & 0690200501 & 2822 \\
  & HD36960 & 4XMM~J053502.6-060007 & 83.761 & -6.002 & 0690200501 & 17546 \\
  & HRC 132 & 4XMM~J053515.0-044442 & 83.813 & -4.745 & 0049560301 & 4549 \\
  & Parenago 1578 & 4XMM~J053440.9-063434 & 83.670 & -6.576 & 0503560701 & 13662 \\
  & UCAC4 427-010050 & 4XMM~J053044.9-044411 & 82.687 & -4.737 & 0763720401 & 3424 \\
  & 2MASS J05343190-0526368 & 4XMM~J053431.5-052639 & 83.632 & -5.444 & 0403200101 & 11919 \\
  & 2MASS J05353534-0511114 & 4XMM~J053535.3-051112 & 83.897 & -5.187 & 0093000101 & 2916 \\
NGC 1333 & ASR 106 & 4XMM~J032859.3+311548 & 52.247 & 31.264 & 0065820101 & 4112 \\
  & EM*~LkHA~270 & 4XMM~J032917.6+312245 & 52.324 & 31.379 & 0056820101 & 2577 \\
  & Gaia DR3 121032079618050560 & 4XMM~J032810.9+311354 & 52.046 & 31.231 & 0065820101 & 2966 \\
Mamajek 1 & EI Cha & 4XMM~J084223.6-790403 & 130.599 & -79.068 & 0605950101 & 10200 \\
  & EL Cha & 4XMM~J084238.8-785443 & 130.662 & -78.912 & 0605950101 & 3403 \\
  & EM Cha & 4XMM~J084307.1-790452 & 130.780 & -79.081 & 0605950101 & 24511 \\
  & EN Cha & 4XMM~J084416.5-785908 & 131.069 & -78.986 & 0605950101 & 5296 \\
  & EO Cha & 4XMM~J084431.9-784631 & 131.132 & -78.775 & 0605950101 & 3611 \\
  & EP Cha & 4XMM~J084701.5-785935 & 131.756 & -78.993 & 0605950101 & 24196 \\
  & EQ Cha & 4XMM~J084756.7-785453 & 131.987 & -78.915 & 0605950101 & 4223 \\
ASCC 16 & HD 35408 & 4XMM~J052436.6+014803 & 81.153 & 1.801 & 0554610101 & 2526 \\
  & HD 287848 & 4XMM~J052454.9+013922 & 81.229 & 1.656 & 0554610101 & 8149 \\
  & 2MASS J05250014+0138252 & 4XMM~J052500.1+013825 & 81.251 & 1.640 & 0554610101 & 2673 \\ 
\end{longtable}

\FloatBarrier
\clearpage

\begin{longtable}{llccccc}
\caption{Stellar parameters of our sample from Gaia DR3}\label{table:gaia} \\
\hline\hline
Star & Gaia DR3 ID & $T_{\rm eff}$[K] & $L_{\rm bol}$[L$_\odot$] & Mass[M$_\odot$] & Radius[R$_\odot$] & Distance[pc] \\ 
\hline
\endfirsthead
\caption{continued.}\\
\hline\hline
Star & Gaia DR3 ID & $T_{\rm eff}$[K] & $L_{\rm bol}$[L$_\odot$] & Mass[M$_\odot$] & Radius[R$_\odot$] & Distance[pc] \\  [2pt] 
\hline
\endhead
\hline
\endfoot
\multicolumn{7}{@{}l@{}}{\parbox{\linewidth}{\footnotesize
\noindent
\tablefoottext{a}{Cluster distances are adopted. 
}\\
\tablefoottext{b}{
Values from \citet{2021AJ....161..147B}.
}\\
\tablefoottext{c}{
This star may not be a cluster member, as its distance differs significantly from the cluster distance.
}
}}
\endlastfoot
CHX~15b & 5201353077910070528 & ... & ... & ... & ... & $199_{-6}^{+5}$\\
Hen~3-545 & 5201226389256838528 & $5197_{-17}^{+20}$ & $4.76_{-0.04}^{+0.04}$ & $1.35_{-0.04}^{+0.04}$ & $2.69_{-0.05}^{+0.05}$ & $184_{-1}^{+1}$ \\
VZ~Cha & 5201353936903777152 & $6123_{-58}^{+120}$ & $1.75_{-0.05}^{+0.04}$ & ... & $1.19_{-0.01}^{+0.01}$ & $191_{-1}^{+1}$\\
V557~Ori & 3209528523708699392 & ... & ... & ... & ... & $388_{-6}^{+6}$ \\
V569~Ori & 3209527943889687936 & ... & ... & ... & ... & $458_{-3}^{+3}$ 
~\tablefootmark{a} \\
V575~Ori & 3209529593157126272 & ... & ... & ... & ... & $289_{-11}^{+10}$ \\
V984~Ori & 3017186663990448512 & $3681_{-4}^{+4}$ & $0.25_{-0.01}^{+0.01}$ & ... & $1.22_{-0.03}^{+0.03}$ & $374_{-2}^{+2}$ \\
V1044~Ori & 3017260709226477056 & $5639_{-59}^{+55}$ & $5.15_{-0.11}^{+0.10}$ & $1.31_{-0.04}^{+0.04}$ & $2.38_{-0.03}^{+0.04}$ & $366_{-15}^{+7}$ \\
V1221~Ori & 3017161237784105600 & ... & ... & ... & ... & $426_{-10}^{+11}$ \\
V1414~Ori & 3017260022031719040 & ... & ... & ... & ... & $390_{-4}^{+4}$ \\
V1417~Ori & 3017253871638555264 & $3881_{-6}^{+15}$ & $0.34_{-0.01}^{+0.01}$ & ... & $1.30_{-0.03}^{+0.03}$ & $373_{-17}^{+32}$\\
V1553~Ori & 3017191543073275008 & $6013_{-15}^{+14}$ & $1.44_{-0.03}^{+0.03}$ & $1.08_{-0.04}^{+0.04}$ & $1.10_{-0.02}^{+0.02}$ & $367_{-3}^{+3}$ \\
V1667~Ori & 3209415239649017856 & ... & ... & ... & ... & $371_{-3}^{+3}$ ~\tablefootmark{b} \\
V1740~Ori & 3209529112120792320 & $4804_{-6}^{+6}$ & $0.92_{-0.01}^{+0.01}$ & ... & $1.38_{-0.03}^{+0.03}$ & $386_{-2}^{+2}$ \\
V1760~Ori & 3017375951801904256 & $4987_{-13}^{+12}$ & $3.17_{-0.55}^{+0.77}$ & $1.06_{-0.06}^{+0.07}$ & $2.29_{-0.22}^{+0.26}$ & $393_{-2}^{+2}$ ~\tablefootmark{b} \\
V1878~Ori & 3209724958330998144 & $5117_{-20}^{+21}$ & $8.95_{-0.14}^{+0.15}$ & $1.57_{-0.05}^{+0.04}$ & $3.81_{-0.08}^{+0.08}$ & $394_{-5}^{+6}$ \\
V1963~Ori & 3017267478094926720 & $5472_{-14}^{+55}$ & $8.34_{-0.15}^{+0.21}$ & $1.55_{-0.05}^{+0.05}$ & $3.20_{-0.07}^{+0.08}$ & $424_{-2}^{+2}$ \\
V2087~Ori & 3017143851756494720 & $5740_{-6}^{+7}$ & $5.00_{-0.08}^{+0.08}$ & $1.28_{-0.04}^{+0.04}$ & $2.26_{-0.05}^{+0.05}$ & $354_{-3}^{+3}$ \\
V2098~Ori & 3017137598284115456 & $4938_{-39}^{+96}$ & $1.20_{-0.01}^{+0.02}$ & ... & $1.52_{-0.01}^{+0.01}$ & $367_{-3}^{+3}$ \\
V2108~Ori & 3017249542311500928 & ... & ... & ... & ... & $370_{-14}^{+14}$ \\ 
AR~Ori & 3209530379132223360 & $5515_{-11}^{+16}$ & $1.67_{-0.03}^{+0.03}$ & $0.94_{-0.04}^{+0.04}$ & $1.41_{-0.03}^{+0.03}$ & $396_{-3}^{+3}$ \\
PR~Ori & 3016976829066516096 & ... & ... & ... & ... & $333_{-10}^{+10}$ \\
WW~Ori & 3017260846665436032 & $5137_{-22}^{+25}$ & $4.58_{-0.08}^{+0.07}$ & $1.32_{-0.14}^{+0.04}$ & $2.72_{-0.06}^{+0.06}$ & $395_{-5}^{+5}$ \\
YZ~Ori & 3209533166569829504 & $5785_{-21}^{+18}$ & ... & ... & ... & $382_{-3}^{+3}$ \\
Brun~213 & 3209422249035572736 & $5567_{-43}^{+18}$ & $3.04_{-0.06}^{+0.05}$ & $1.10_{-0.04}^{+0.04}$ & $1.88_{-0.04}^{+0.04}$ & $405_{-2}^{+10}$ \\
Brun~302 & 3017269947698014336 & $5133_{-16}^{+41}$ & $4.05_{-0.05}^{+0.06}$ & $1.26_{-0.05}^{+0.04}$ & $2.54_{-0.05}^{+0.05}$ & $389_{-8}^{+22}$ \\
Brun~311 & 3017267306296233216 & $6603_{-348}^{+50}$ & $9.09_{-0.48}^{+0.62}$ & $1.61_{-0.05}^{+0.05}$ & $2.06_{-0.03}^{+0.06}$ & $497_{-8}^{+9}$ \\
Brun~394 & 3017188893075487104 & ... & ... & ... & ... & $385_{-9}^{+8}$ \\
Brun~656 & 3209527291055512064 & ... & ... & ... & ... & $497_{-12}^{+12}$ \\
Brun~684 & 3017191783588975744 & ... & ... & ... & ... & $330_{-75}^{+54}$ \\
BD-06~1236 & 3017145290567504640 & ... & ... & ... & ... & $370_{-4}^{+4}$ \\
HD~36959 & 3017187866581306368 & $19013_{-166}^{+327}$ & $2690_{-210}^{+250}$ & $6.92_{-0.14}^{+0.18}$ & $4.75_{-0.16}^{+0.18}$ & $376_{-22}^{+28}$ \\
HD~36960 & 3017187866581304832 & $23082_{-42}^{+48}$ & ... & ... & ... & $419_{-20}^{+17}$ \\
HRC~132 & 3209578143467255680 & $5516_{-8}^{+42}$ & $11.1_{-0.4}^{+0.5}$ & $1.68_{-0.05}^{+0.05}$ & $3.63_{-0.10}^{+0.10}$ & $423_{-1}^{+3}$ \\
Parenago~1578 & 3017130279659878400 & $5198_{-31}^{+38}$ & $2.81_{-0.04}^{+0.05}$ & $1.08_{-0.05}^{+0.04}$ & $2.09_{-0.04}^{+0.05}$ & $406_{-9}^{+10}$ \\
UCAC4~427-010050 & 3209709569465229440 & $4212_{-22}^{+4}$ & $0.21_{-0.01}^{+0.01}$ & ... & $0.86_{-0.02}^{+0.02}$ & $127_{-0}^{+0}$ ~\tablefootmark{c} \\
2MASS~J05343190-0526368 & 3017270329953200128 & ... & ... & ... & ... & $366_{-8}^{+8}$ \\ 
2MASS~J05353534-0511114 & 3209527772090999040 & ... & ... & ... & ... & $383_{-8}^{+8}$ \\ 
ASR~106 & ... & ... & ... & ... & ... & $324_{-3}^{+3}$ ~\tablefootmark{a} \\ 
EM*~LkHA~270 & 121406699550262528 & ... & ... & ... & ... & $324_{-3}^{+3}$ ~\tablefootmark{a} \\
Gaia~DR3~121032079618050560 & 121032079618050560 & $3425_{-2}^{+2}$ & $0.08_{-0.01}^{+0.01}$ & ... & $0.80_{-0.03}^{+0.03}$ & $219_{-1}^{+1}$ \\
EI~Cha & 5209019521517870080 & $3861_{-12}^{+3}$ & $0.32_{-0.01}^{+0.01}$ & ... & $1.26_{-0.03}^{+0.03}$ & $99_{-0}^{+3}$ \\
EL~Cha & 5209038423669475712 & $3529_{-3}^{+3}$ & $0.14_{-0.01}^{+0.01}$ & ... & $1.01_{-0.03}^{+0.03}$ & $98_{-0}^{+0}$ \\
EM~Cha & 5209024576694905472 & $5506_{-3}^{+4}$ & $1.76_{-0.01}^{+0.01}$ & $0.95_{-0.04}^{+0.04}$ & $1.46_{-0.03}^{+0.03}$ & $99_{-0}^{+0}$ \\
EN~Cha & 5209025366968905856 & ... & ... & ... & ... & $112_{-1}^{+1}$ ~\tablefootmark{a} \\
EO~Cha & 5209135352491538432 & $4116_{-2}^{+2}$ & $0.33_{-0.01}^{+0.01}$ & ... & $1.12_{-0.02}^{+0.02}$ & $98_{-0}^{+0}$ \\
EP~Cha & 5209023133585929984 & $4532_{-8}^{+11}$ & $0.71_{-0.01}^{+0.01}$ & $0.85_{-0.04}^{+0.04}$ & $1.37_{-0.03}^{+0.03}$ & $99_{-0}^{+0}$ \\
EQ~Cha & 5209118305765620096 & ... & ... & ... & ... & $97_{-0}^{+0}$ \\
HD~35408 & 3222275887858172288 & ... & ... & ... & ... & $337_{-4}^{+4}$ \\
HD~287848 & 3222258845427956352 & $5798_{-18}^{+16}$ & $3.64_{-0.04}^{+0.04}$ & $1.16_{-0.04}^{+0.04}$ & $1.89_{-0.04}^{+0.04}$ & $341_{-2}^{+2}$ \\
2MASS~J05250014+0138252 & 3222258772412327296 & $5061_{-23}^{+17}$ & $0.88_{-0.01}^{+0.01}$ & $0.93_{-0.04}^{+0.04}$ & $1.23_{-0.03}^{+0.03}$ & $374_{-4}^{+16}$ \\
\hline
\end{longtable}

\onecolumn
\section{Analysis results}
This appendix presents the additional X-ray light curves and the tables summarizing the timing and spectral analysis results.

\begin{figure*}[ht!]
    \includegraphics[width=0.33\textwidth]{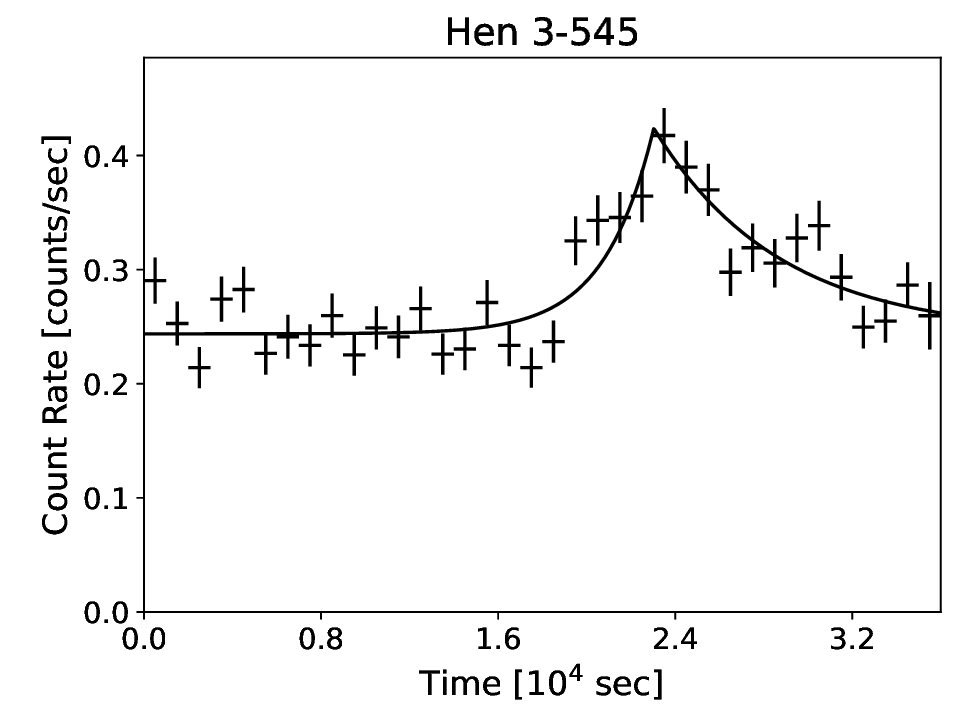} \hfill
    \includegraphics[width=0.33\textwidth]{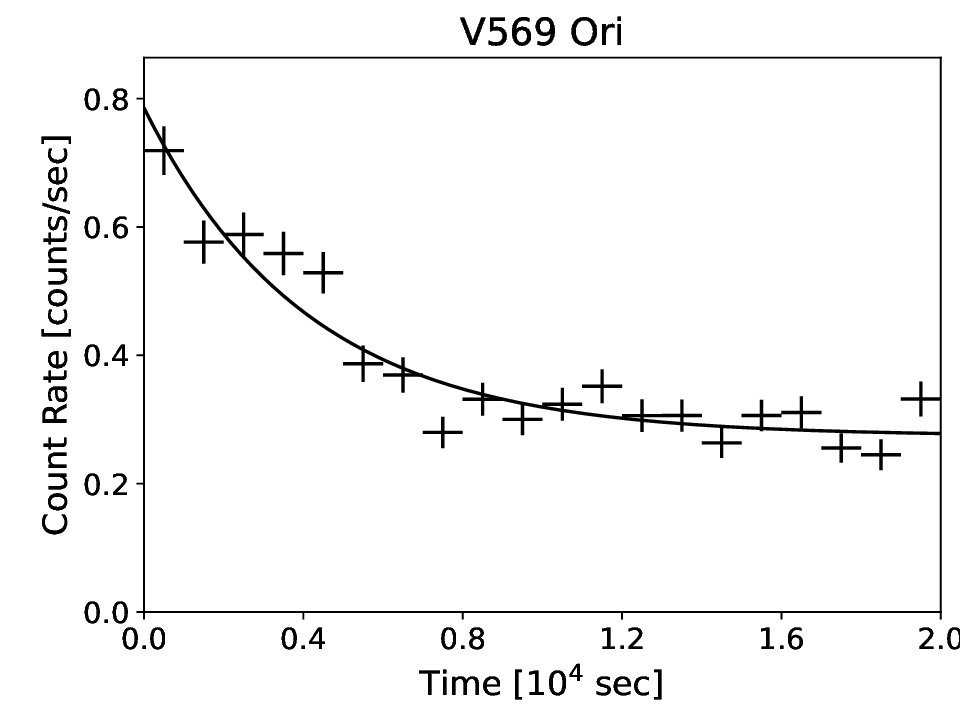} \hfill
    \includegraphics[width=0.33\textwidth]{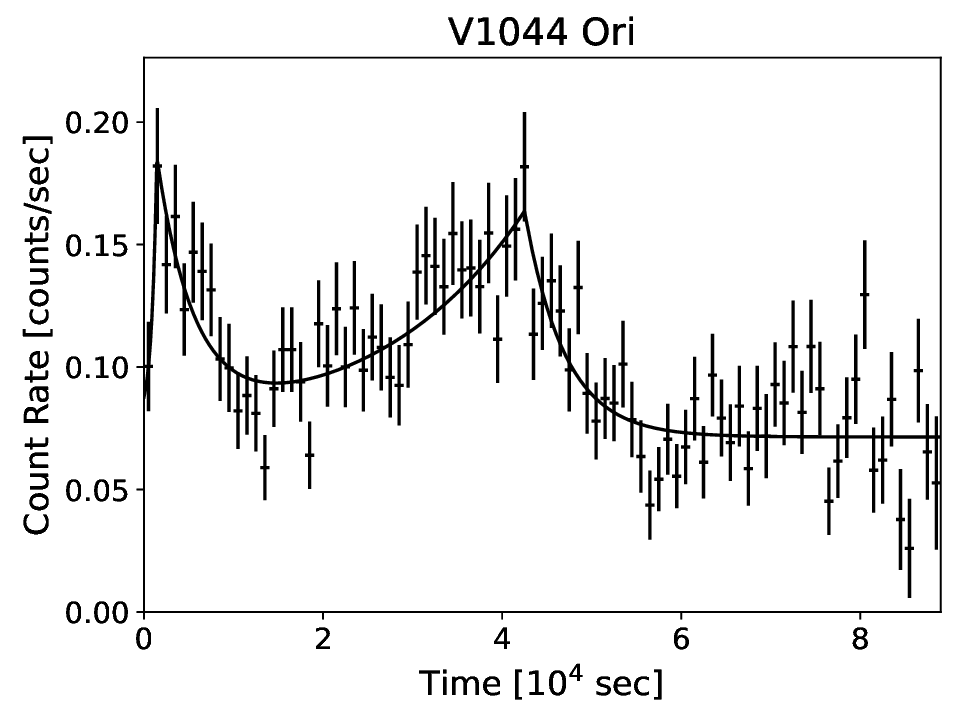} \vspace{3mm}
    \includegraphics[width=0.33\textwidth]{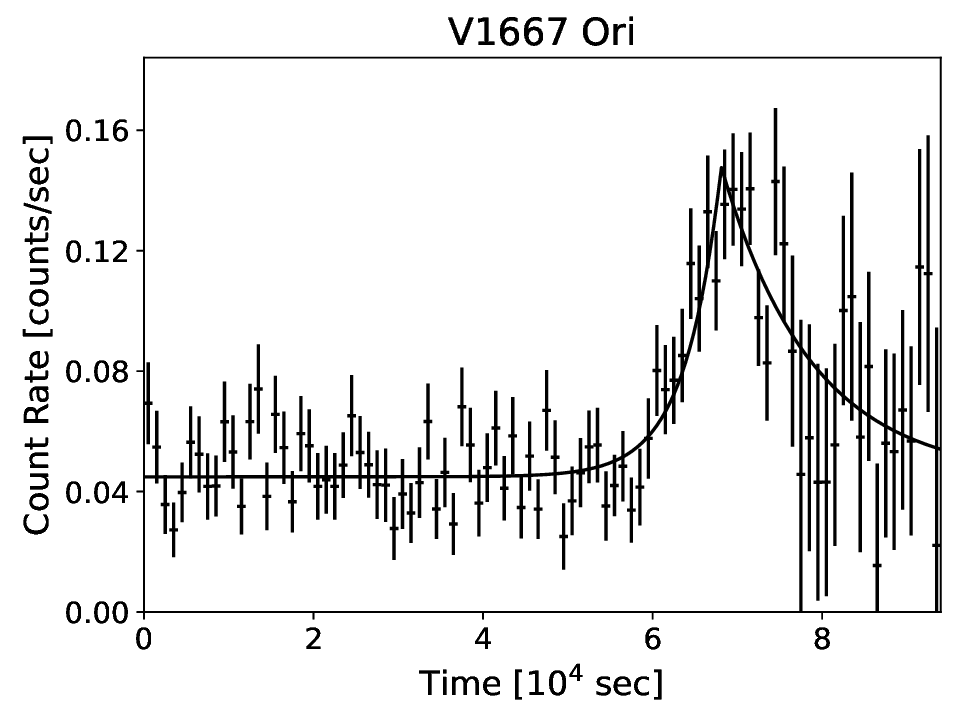} \hfill
    \includegraphics[width=0.33\textwidth]{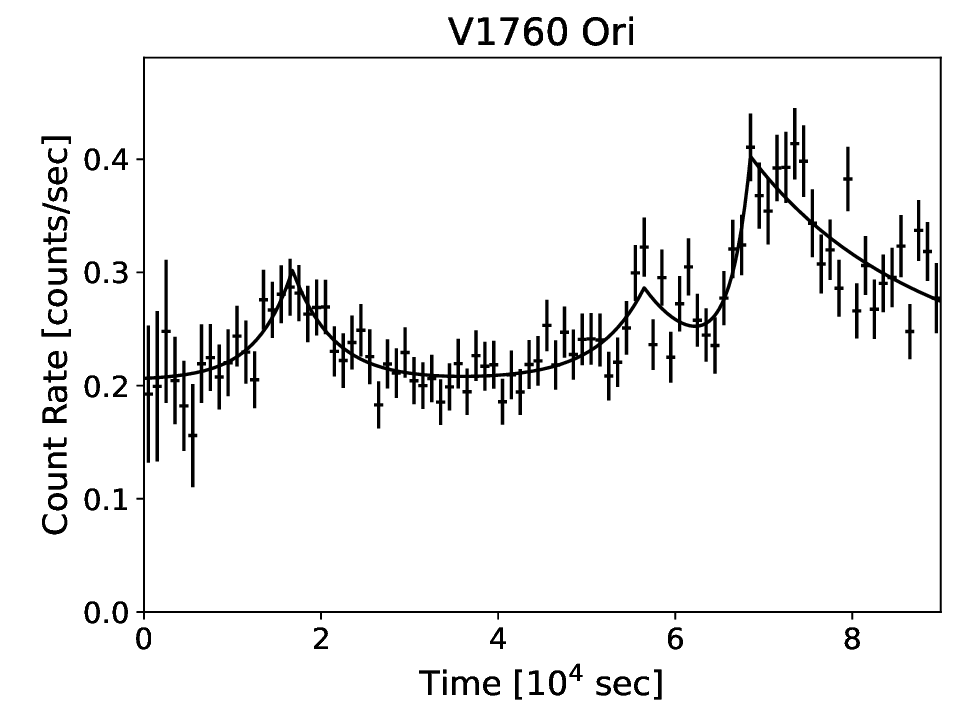} \hfill
    \includegraphics[width=0.33\textwidth]{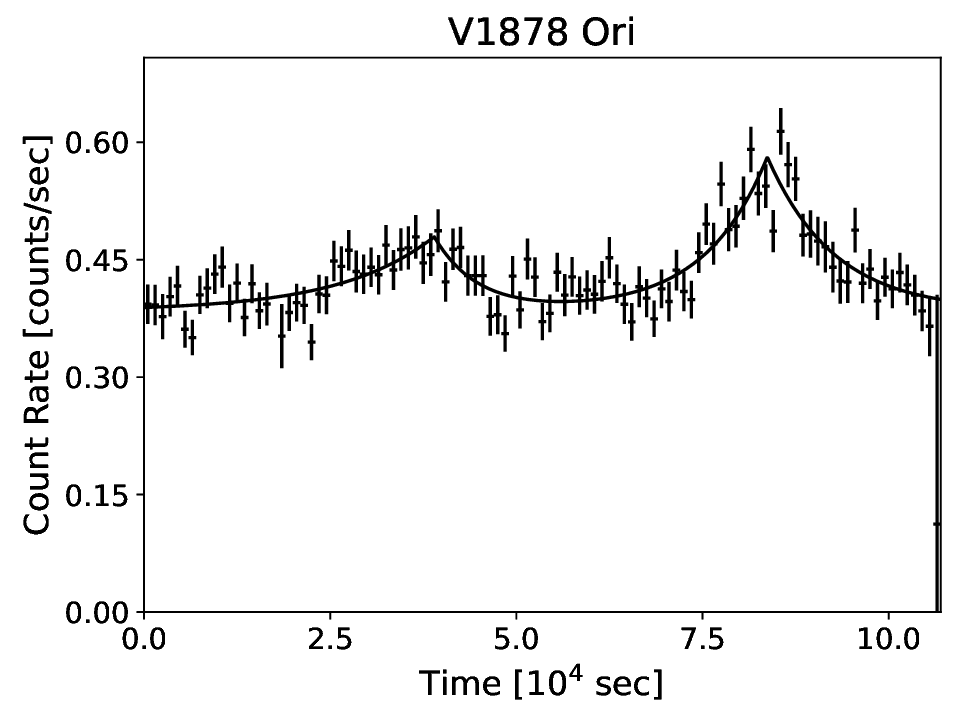} \vspace{3mm}
    \includegraphics[width=0.33\textwidth]{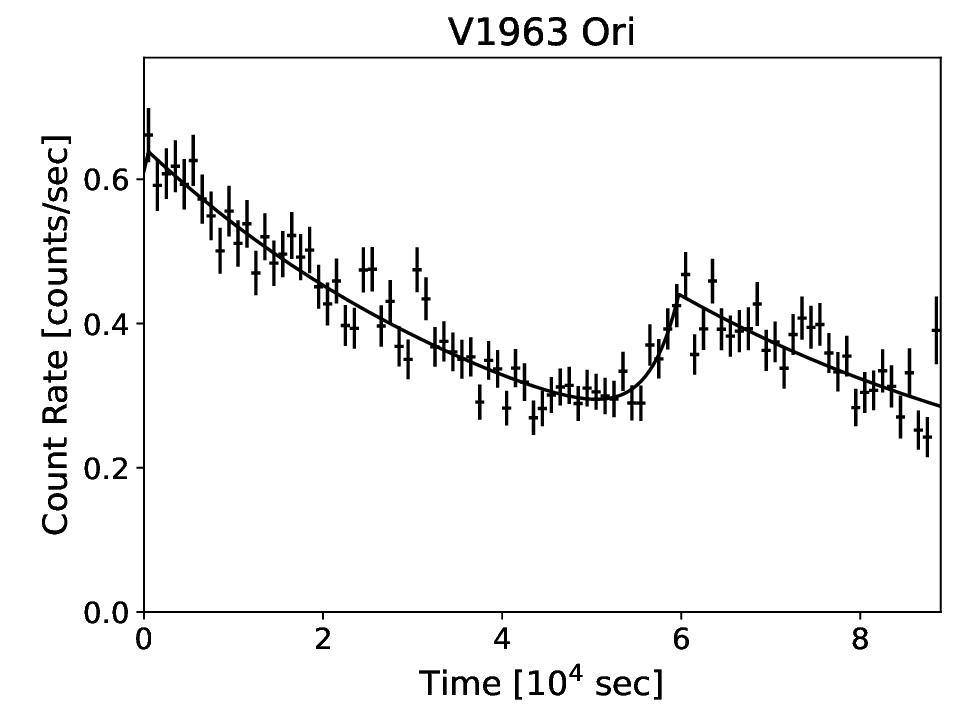} \hfill
    \includegraphics[width=0.33\textwidth]{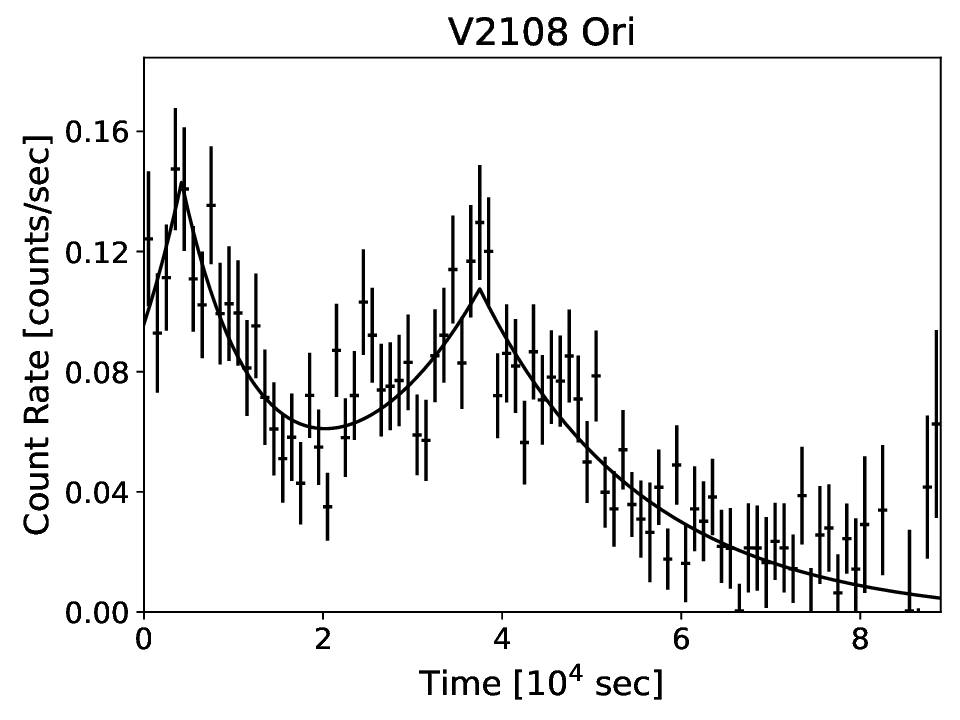} \hfill
    \includegraphics[width=0.33\textwidth]{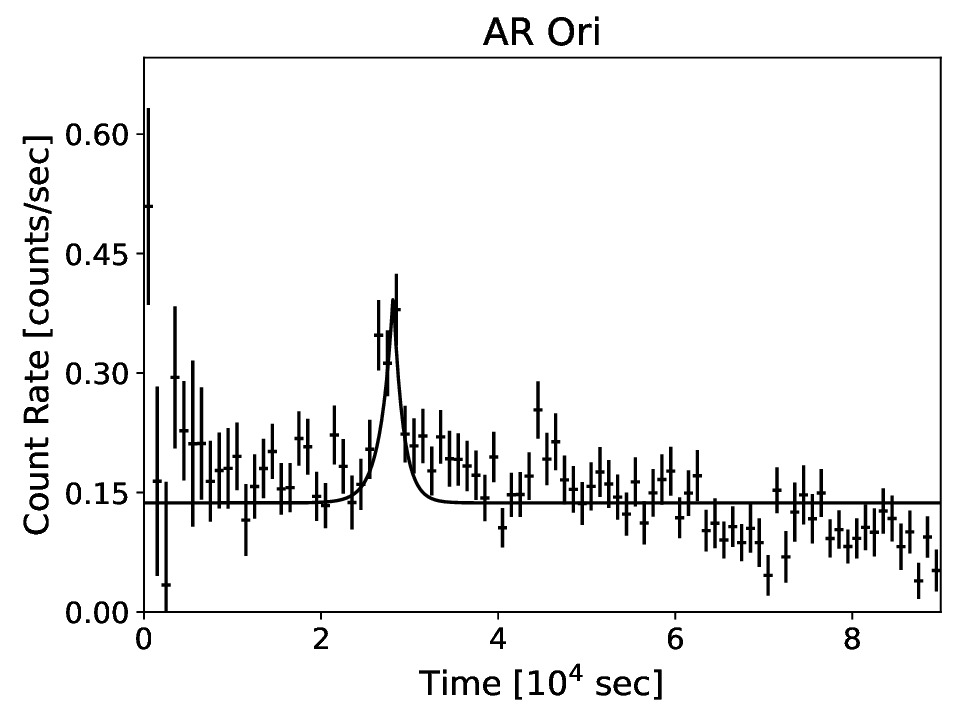} \vspace{3mm}
    \includegraphics[width=0.33\textwidth]{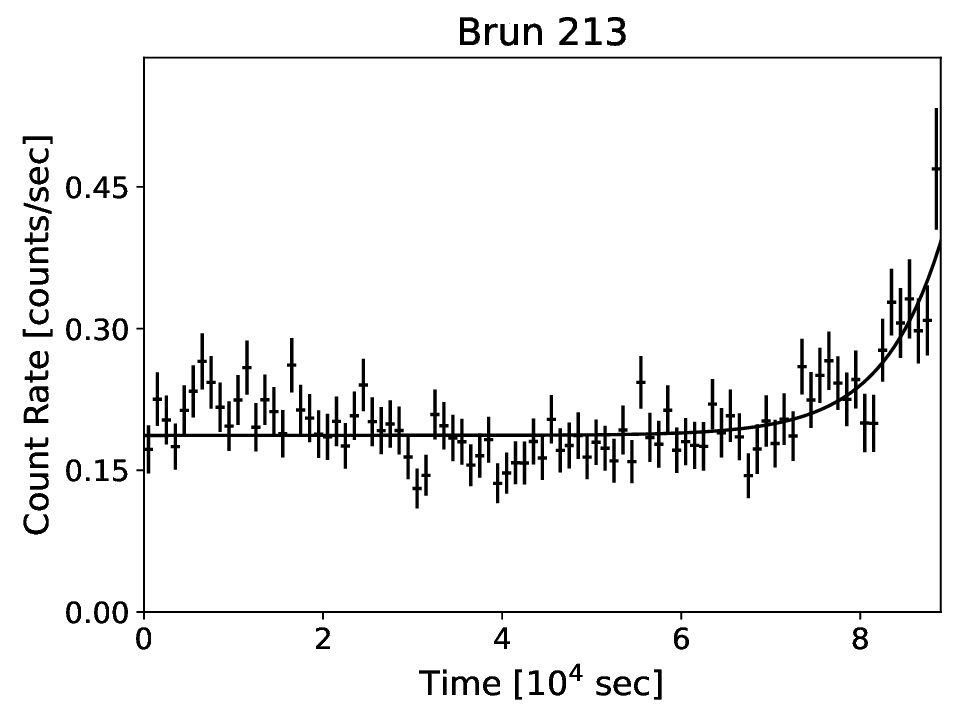} \hfill
    \includegraphics[width=0.33\textwidth]{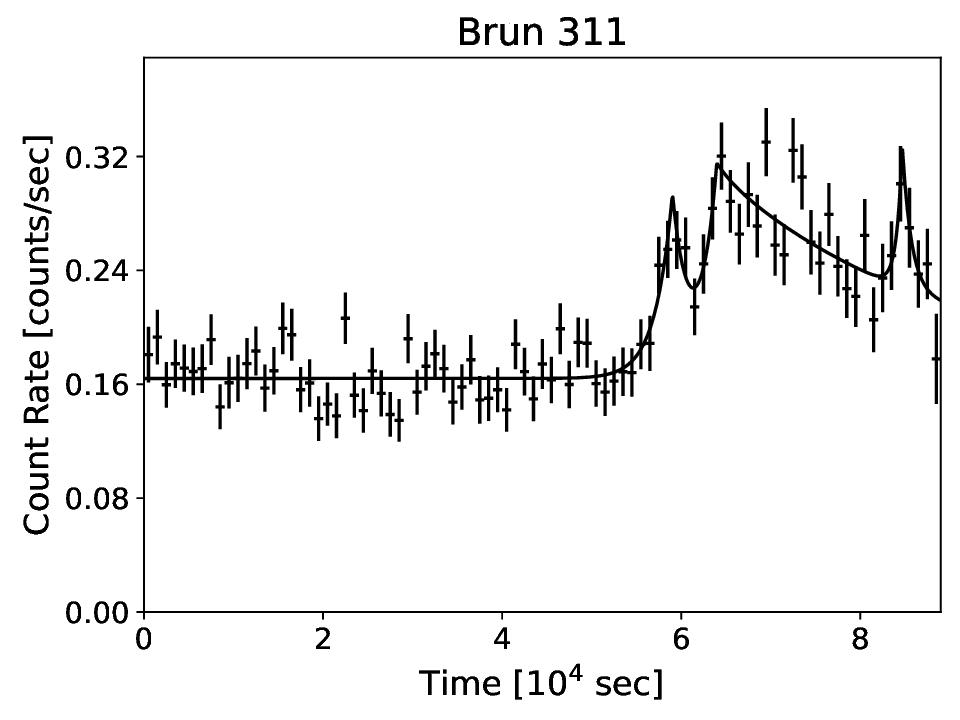} \hfill
    \includegraphics[width=0.33\textwidth]{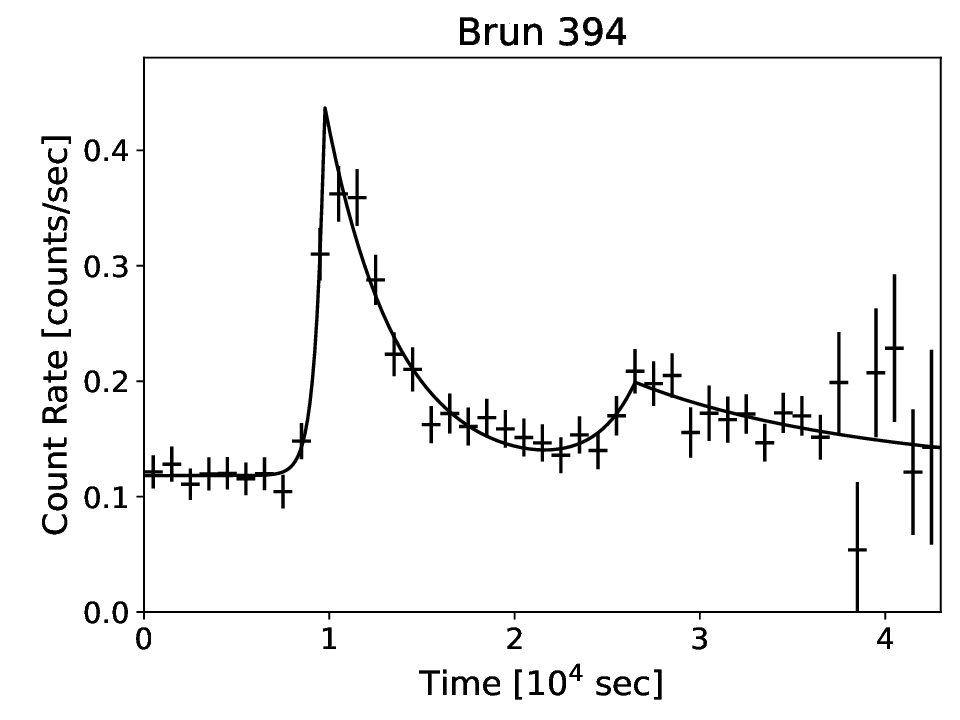} \vspace{3mm} 
    \caption{Additional X-ray light curves of our sample that exhibited outbursts during the observations, overlaid with the best-fit light-curve models. The data points represent the observed count rates, while the solid lines indicate the model fits.}
    \label{fig:lc_image_a}
\end{figure*}
\addtocounter{figure}{-1}
\begin{figure*}[ht!]
    \includegraphics[width=0.33\textwidth]{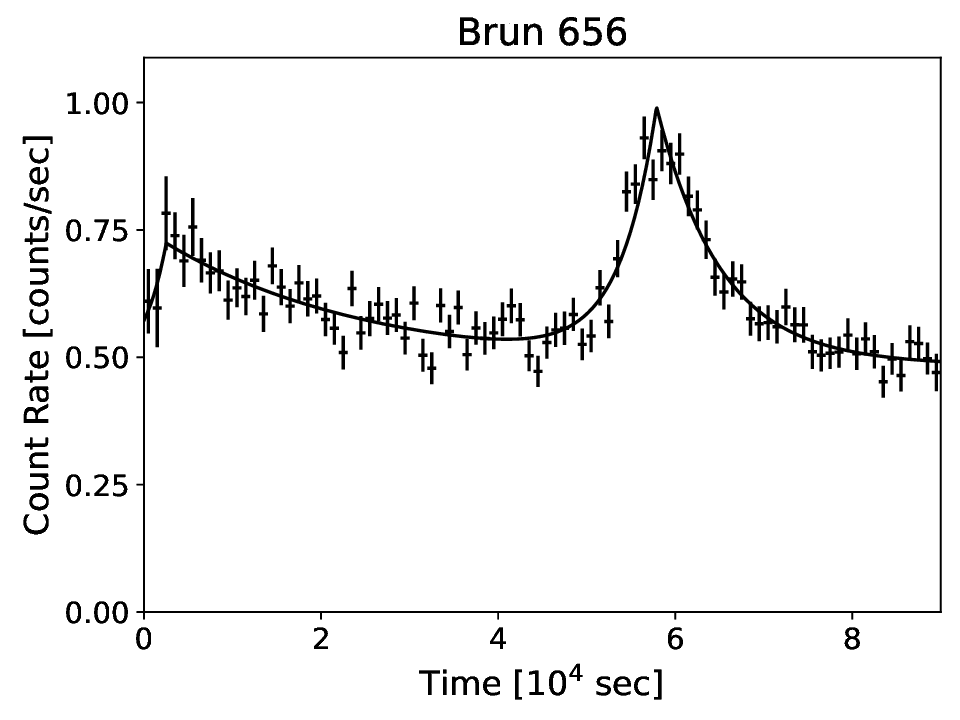} \hfill
    \includegraphics[width=0.33\textwidth]{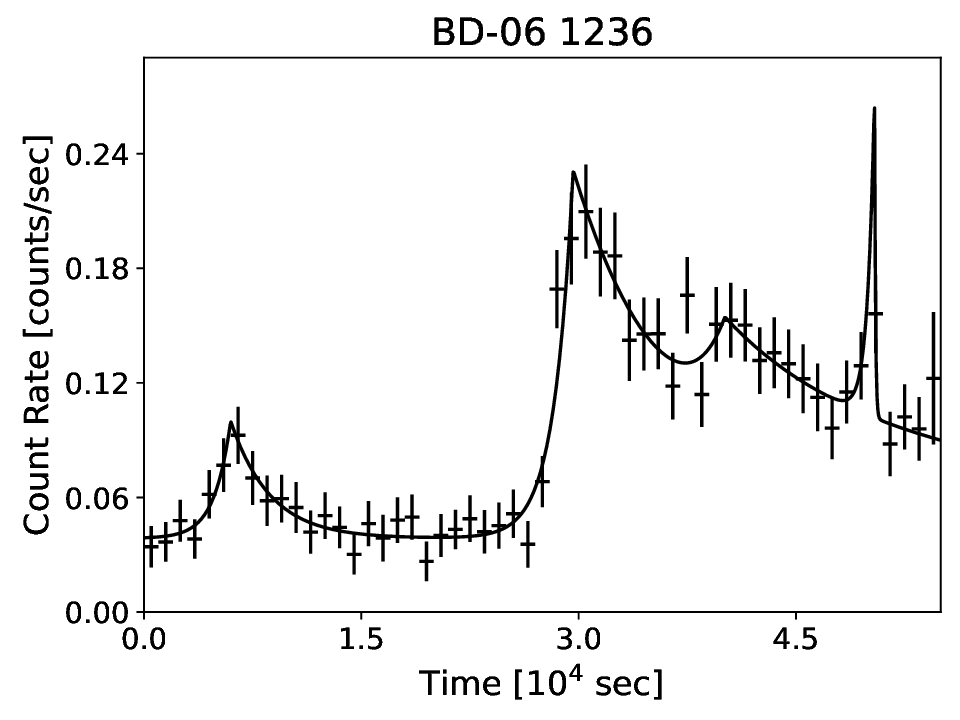} \hfill
    \includegraphics[width=0.33\textwidth]{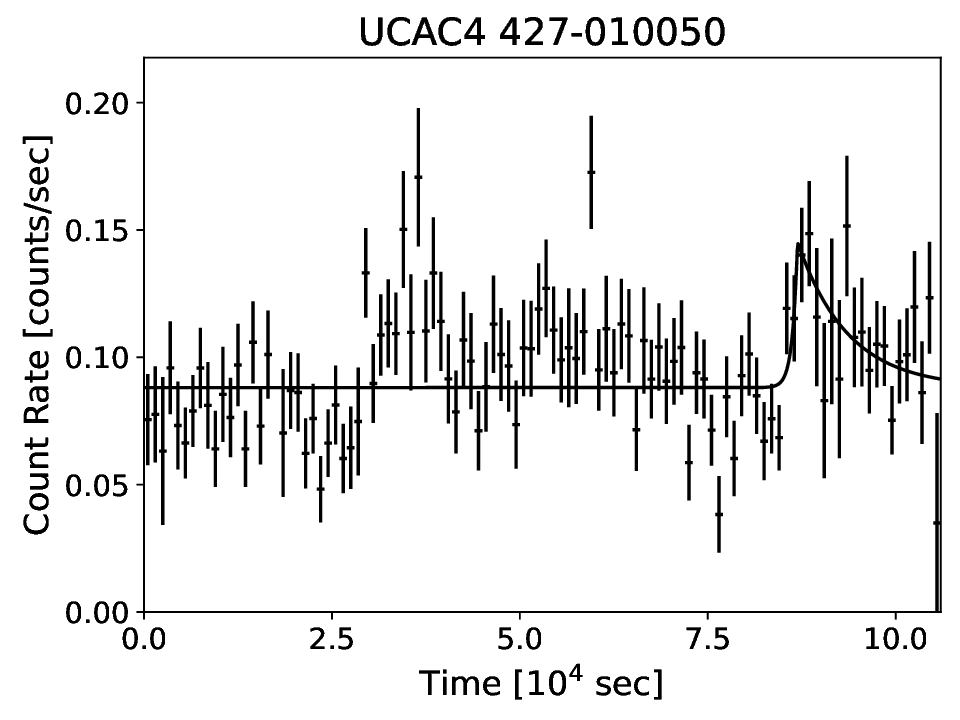} \vspace{3mm}
    \includegraphics[width=0.33\textwidth]{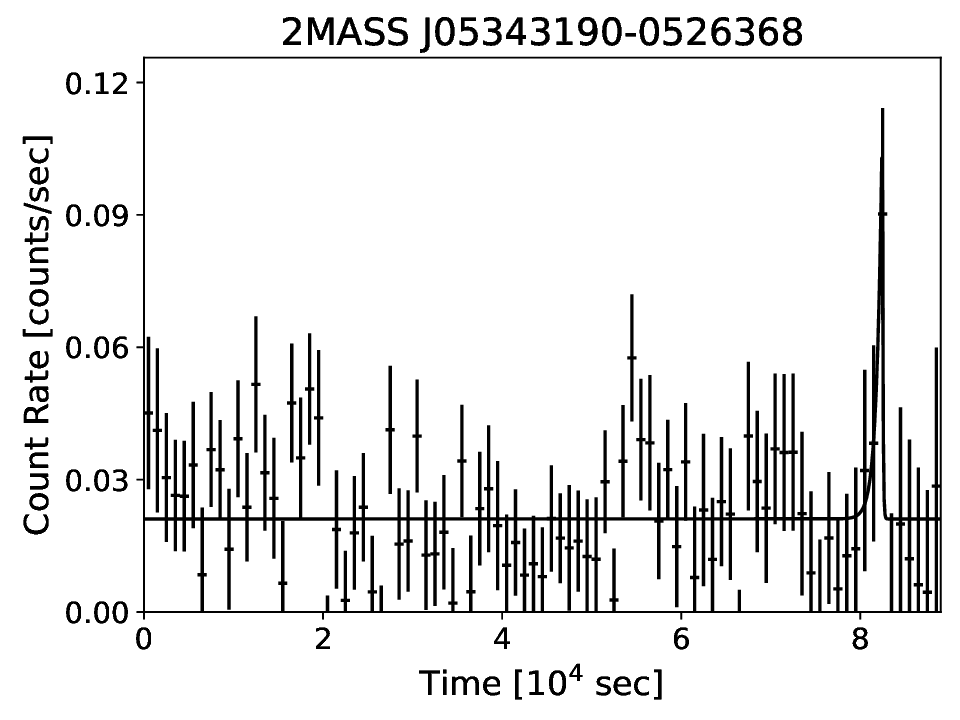} \hfill
    \includegraphics[width=0.33\textwidth]{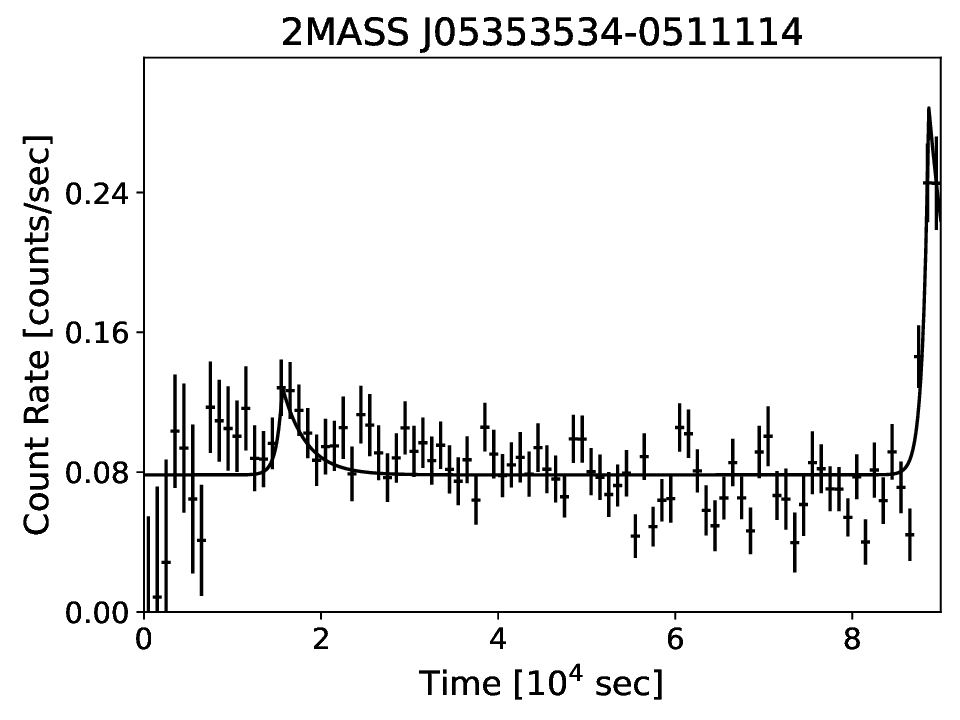} \hfill
    \includegraphics[width=0.33\textwidth]{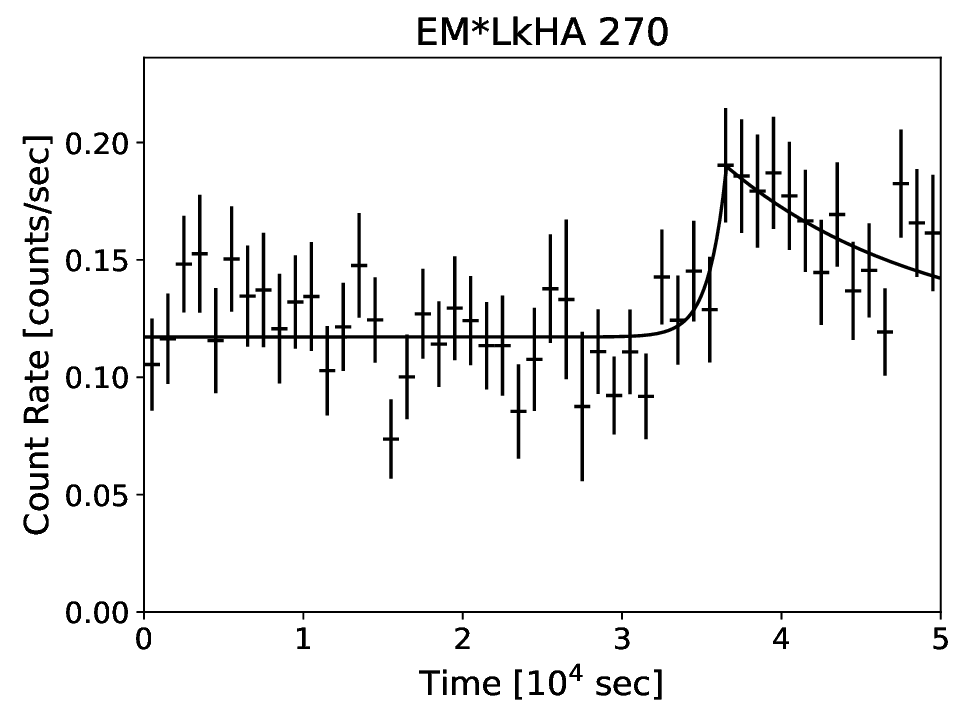} \vspace{3mm}
    \includegraphics[width=0.33\textwidth]{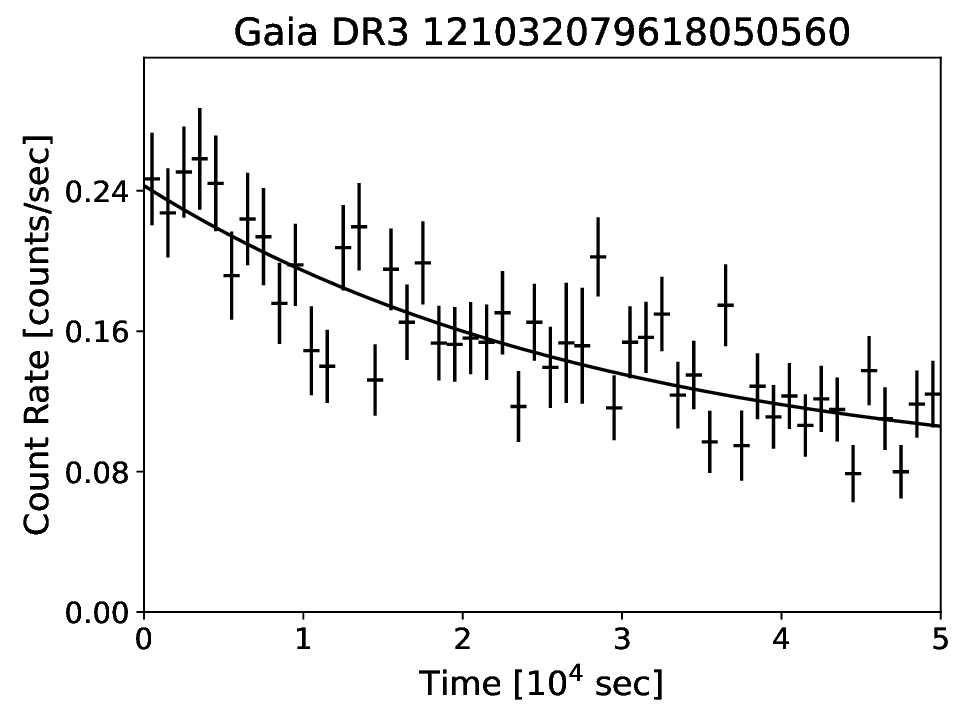} \hfill
    \includegraphics[width=0.33\textwidth]{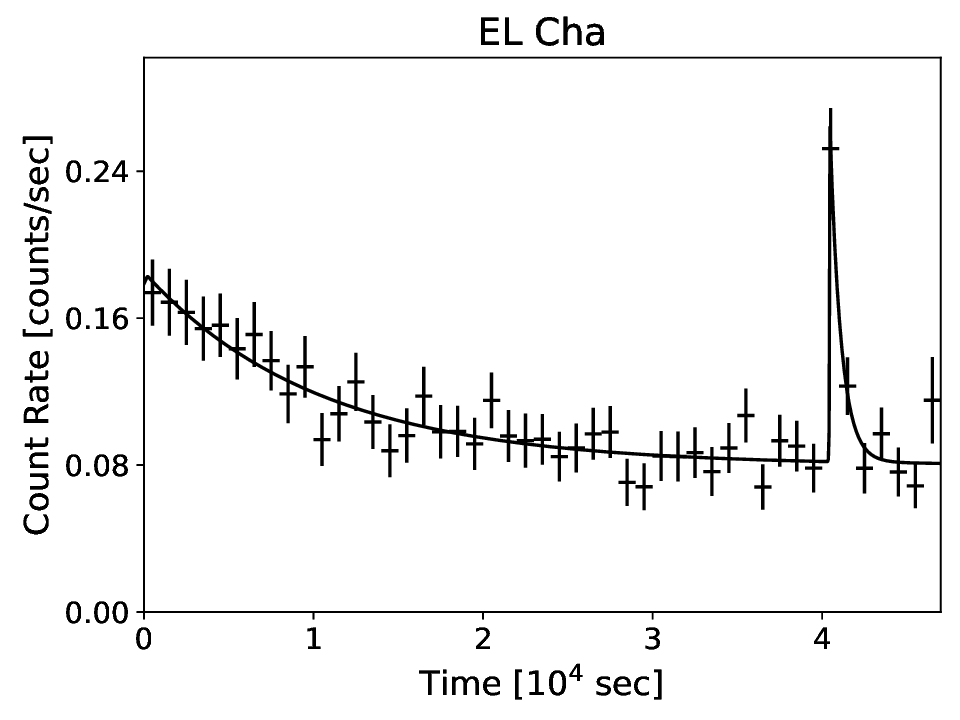} \hfill
    \includegraphics[width=0.33\textwidth]{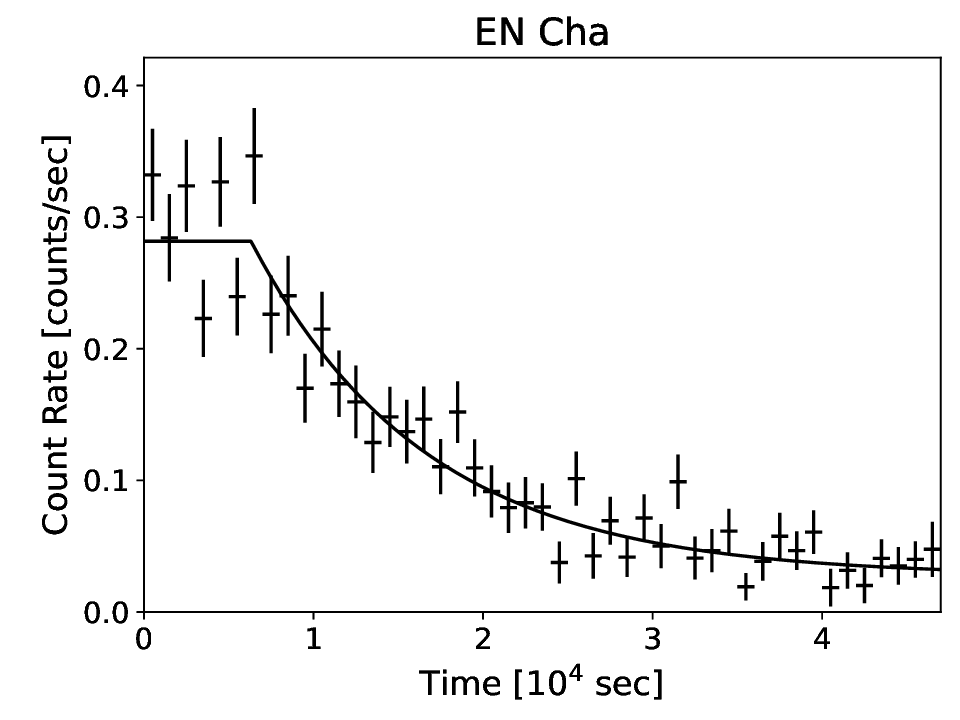} \vspace{3mm}
    \includegraphics[width=0.33\textwidth]{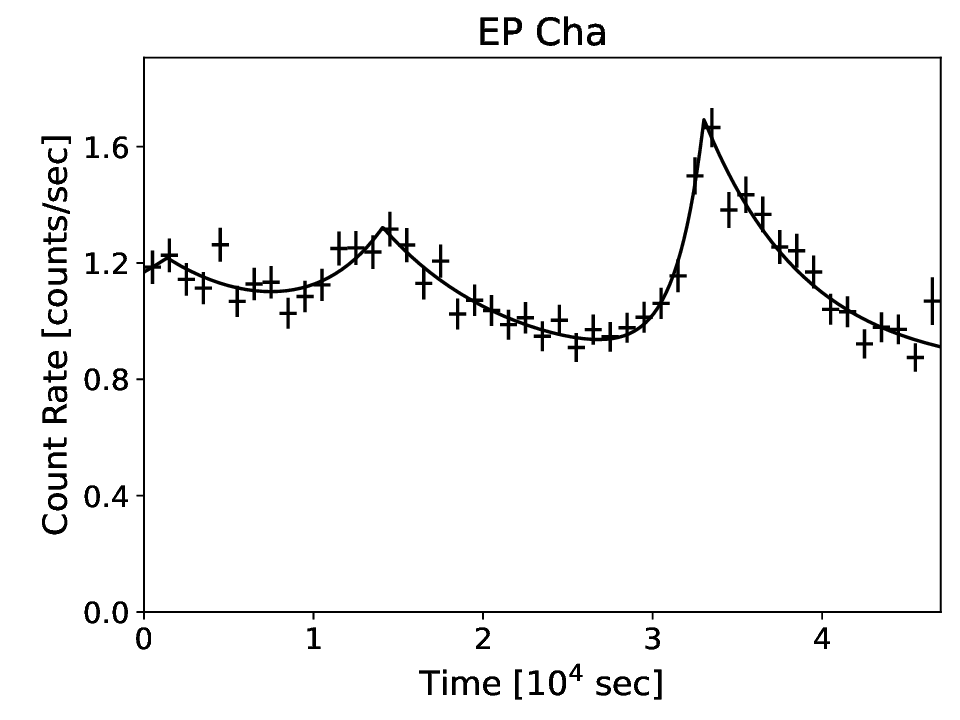} \hfill
    \includegraphics[width=0.33\textwidth]{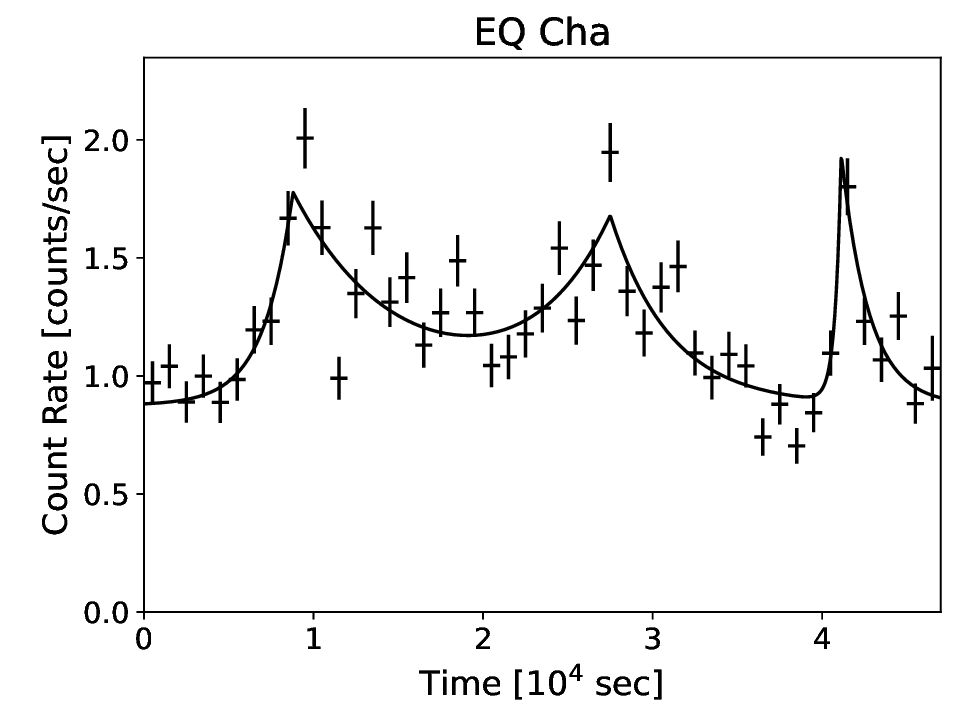} \hfill
    \includegraphics[width=0.33\textwidth]{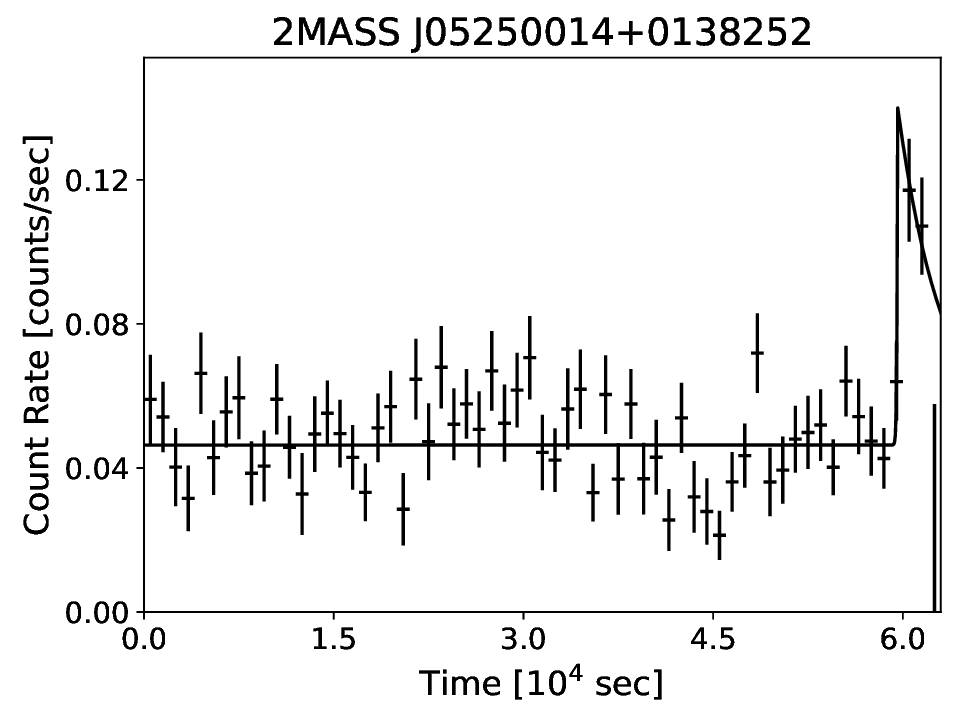} \vspace{3mm} 
    \caption{continued.}
\end{figure*}

\FloatBarrier
\clearpage

\begin{longtable}{llccc}
\caption{Timing analysis results}\label{table:lc} \\
\hline\hline
Star & Phase & Count rate [$\mathrm{count\,s^{-1}}$]~\tablefootmark{a} & Rise time [${\rm s}$]~\tablefootmark{a} & Decay time [${\rm s}$]~\tablefootmark{a} \\ 
\hline
\endfirsthead
\caption{continued.}\\
\hline\hline
Star & Phase & Count rate [$\mathrm{count\,s^{-1}}$] & Rise time [${\rm s}$] & Decay time [${\rm s}$] \\  
\hline
\endhead
\hline
\endfoot
\multicolumn{5}{@{}l@{}}{\parbox{\linewidth}{\footnotesize
\noindent
\tablefoottext{a}{When parameters could not be constrained, best-fit values are listed where available; otherwise, entries are left blank.}
}}
\endlastfoot
CHX 15b & Quiescent & 0.15 $\pm$ 0.00  & ... & ... \\
 & Outburst & 0.39 $\pm$ 0.05  & 1200 $\pm$ 200  & 1900 $\pm$ 400 \\ 
Hen 3-545 & Quiescent & 0.24 $\pm$ 0.01  & ... & ... \\ 
 & Outburst & 0.18 $\pm$ 0.02  & 2200 $\pm$ 600  & 5700 $\pm$ 1000 \\ 
VZ Cha & Quiescent & 0.038 $\pm$ 0.002  & ... & ... \\
 & Outburst 1 & 0.23 $\pm$ 0.02  & 2500 $\pm$ 500  & 8800 $\pm$ 900 \\ 
 & Outburst 2 & 0.40 $\pm$ 0.02  & 1400 $\pm$ 300  & 20000 $\pm$ 1000 \\ 
 & Outburst 3 & 0.11 $\pm$ 0.02  & 3200 $\pm$ 1700  & 4900 $\pm$ 1800 \\
V557 Ori & Quiescent & $0.070 \pm 0.001$ & ... & ... \\ 
V569 Ori & Quiescent & 0.28 $\pm$ 0.01  & ... & ... \\ 
 & Outburst & 0.46 & 2000 & 4000$\pm$ 600 \\ 
V575 Ori & Quiescent & 0.26 $\pm$ 0.01  & ... & ... \\ 
V984 Ori & Quiescent & 0.21 $\pm$ 0.00  & ... & ... \\ 
V1044 Ori & Quiescent & 0.071 $\pm$ 0.003  & ... & ... \\ 
 & Outburst 1 & 0.11 & 580 & 4100$\pm$ 1200 \\ 
 & Outburst 2 & 0.091 $\pm$ 0.010  & 17000 $\pm$ 4000  & 4600 $\pm$ 1300 \\ 
V1221 Ori & Quiescent & 0.18 $\pm$ 0.00  & ... & ... \\ 
V1414 Ori & Quiescent & 0.35 $\pm$ 0.00  & ... & ... \\
 & Outburst 1 & 5.4 $\pm$ 0.1  & 1200 $\pm$ 0  & 5800 $\pm$ 100 \\
 & Outburst 2 & 0.093 $\pm$ 0.021  & 2500 $\pm$ 1700  & 5600 $\pm$ 2700 \\
V1417 Ori & Quiescent & 0.17 & ... & ... \\ 
 & Outburst 1 & 1.3 & 1300 & 9600 \\ 
 & Outburst 2 & 7.5 & 8500 & ... \\ 
V1553 Ori & Quiescent & $1.3 \pm 0.0$ & ... & ... \\ 
V1667 Ori & Quiescent & 0.045 $\pm$ 0.002  & ... & ... \\ 
 & Outburst & 0.10 $\pm$ 0.01  & 4200 $\pm$ 900  & 11000 $\pm$ 3000 \\ 
V1740 Ori & Quiescent & 0.066 $\pm$ 0.002  & ... & ... \\ 
V1760 Ori & Quiescent & 0.21 $\pm$ 0.01  & ... & ... \\
 & Outburst 1 & 0.097 $\pm$ 0.018  & 4000 $\pm$ 2000  & 4600 $\pm$ 2300 \\
 & Outburst 2 & 0.081 $\pm$ 0.019  & 5300 $\pm$ 2700  & 7100 $\pm$ 8100 \\
 & Outburst 3 & 0.18 $\pm$ 0.03  & 2300 $\pm$ 1100  & 22000 $\pm$ 5400 \\
V1878 Ori & Quiescent & 0.38 $\pm$ 0.01  & ... & ... \\ 
 & Outburst 1 & 0.094 $\pm$ 0.013  & 13000 $\pm$ 5000  & 5700 $\pm$ 2600 \\
 & Outburst 2 & 0.20 $\pm$ 0.02  & 8400 $\pm$ 1700  & 9000 $\pm$ 1900 \\
V1963 Ori & Quiescent & 0.093 $\pm$ 0.131  & ... & ... \\ 
 & Outburst 1 & 0.545 & ... & 47000$\pm$ 18000 \\ 
 & Outburst 2 & 0.19 $\pm$ 0.02  & 3500 $\pm$ 1200  & 51000 $\pm$ 35000 \\ 
V2087 Ori & Quiescent & $0.43 \pm 0.00$ & ... & ... \\ 
V2098 Ori & Quiescent & 0.091 $\pm$ 0.002  & ... & ... \\ 
V2108 Ori & Quiescent & $\leq 0.0081$ & ... & ... \\ 
 & Outburst 1 & 0.55 $\pm$ 0.08  & 9900 $\pm$ 9400  & 8500 $\pm$ 3200 \\ 
 & Outburst 2 & 0.47 $\pm$ 0.04  & 19000 $\pm$ 9000  & 18000 $\pm$ 5000 \\ 
AR Ori & Quiescent & 0.14 $\pm$ 0.00  & ... & ... \\
 & Outburst & 0.26 $\pm$ 0.07  & 1400 $\pm$ 600  & 1100 $\pm$ 500 \\
PR Ori & Quiescent & 0.37 $\pm$ 0.00  & ...  & ...  \\
WW Ori & Quiescent & 0.11 $\pm$ 0.00  & ...  & ...  \\ 
YZ Ori & Quiescent & 0.091 $\pm$ 0.002  & ...  & ...  \\
Brun 213 & Quiescent & 0.19 & ...  & ...  \\
 & Outburst & 0.25 & 6600 & ...  \\ 
Brun 302 & Quiescent & 0.14 $\pm$ 0.00  & ...  & ...  \\ 
Brun 311 & Quiescent & 0.16 $\pm$ 0.00  & ...  & ...  \\
 & Outburst & 0.14 $\pm$ 0.01  & 5800 $\pm$ 1100  & 33000 $\pm$ 8000 \\ 
Brun 394 & Quiescent & 0.12 $\pm$ 0.01  & ...  & ...  \\ 
 & Outburst 1 & 0.32 $\pm$ 0.03  & 490 $\pm$ 160  & 3800 $\pm$ 400 \\ 
 & Outburst 2 & 0.077 $\pm$ 0.015  & 2600 $\pm$ 900  & 14000 $\pm$ 7000 \\ 
Brun 656 & Quiescent & 0.51 $\pm$ 0.02  & ...  & ...  \\ 
 & Outburst 1 & 0.26 $\pm$ 0.04  & 3000  & 24000 $\pm$ 5000 \\
 & Outburst 2 & 0.52 $\pm$ 0.03  & 4200 $\pm$ 500  & 7300 $\pm$ 900 \\ 
Brun 684 & Quiescent & 0.33 $\pm$ 0.01  & ...  & ...  \\ 
BD-06 1236 & Quiescent & 0.040 $\pm$ 0.003  & ...  & ...  \\ 
 & Outburst 1 & 0.062 $\pm$ 0.018  & 1000 $\pm$ 700  & 2400 $\pm$ 1100 \\
 & Outburst 2 & 0.15 $\pm$ 0.01  & 610 $\pm$ 200  & 28000 $\pm$ 5000 \\ 
HD 36959 & Quiescent & 0.086 $\pm$ 0.003  & ...  & ...  \\ 
HD 36960 & Quiescent & ...  & ...  & ...  \\ 
HRC 132 & Quiescent & 0.43 $\pm$ 0.01  & ...  & ...  \\ 
Parenago 1578 & Quiescent & $0.83 \pm 0.01$  & ...  & ...  \\ 
UCAC4 427-010050 & Quiescent & 0.088 $\pm$ 0.002  & ...  & ...  \\
 & Outburst & 0.057 $\pm$ 0.016  & 680 & 6700 $\pm$ 2700 \\
2MASS J05343190-0526368 & Quiescent & 0.021 $\pm$ 0.002  & ...  & ...  \\ 
 & Outburst & 0.093 & 600 & 64 \\ 
2MASS J05353534-0511114 & Quiescent & 0.078 $\pm$ 0.002  & ...  & ...  \\
 & Outburst 1 & 0.050 $\pm$ 0.022  & 650 $\pm$ 690  & 2700 $\pm$ 1800 \\ 
 & Outburst 2 & 0.21 & 670 $\pm$ 190  & 3500 \\ 
ASR 106 & Quiescent & 0.041 $\pm$ 0.002  & ...  & ...  \\ 
 & Outburst & 0.90 $\pm$ 0.07  & 2500 $\pm$ 100  & 7800 $\pm$ 4300 \\ 
EM*~LkHA~270 & Quiescent & 0.12 $\pm$ 0.00  & ...  & ...  \\ 
 & Outburst & 0.073 $\pm$ 0.017  & 1100 $\pm$ 800  & 13000 $\pm$ 7000 \\
Gaia DR3 121032079618050560 & Quiescent & 0.076 & ...  & ...  \\ 
 & Outburst & ... & ...  & 29000 \\ 
EI Cha & Quiescent & 0.242 & ...  & ...  \\
 & Outburst 1 & 0.10 & 420 & 4500  \\
 & Outburst 2 & 0.72 & 3000 & ...  \\ 
EL Cha & Quiescent & 0.080 & ... & ...  \\
 & Outburst 1 & ...  & ...  & 10000 \\
 & Outburst 2 & 0.18 & 18 & 650 \\ 
EM Cha & Quiescent & 0.59 $\pm$ 0.01  & ...  & ...  \\ 
EN Cha & Quiescent & 0.027 $\pm$ 0.007  & ...  & ...  \\ 
 & Outburst & 0.25 $\pm$ 0.03  & - & 10000 $\pm$ 1000 \\ 
EO Cha & Quiescent & 0.19 $\pm$ 0.00  & ...  & ...  \\ 
EP Cha & Quiescent & 0.83 $\pm$ 0.08  & ...  & ...  \\
 & Outburst 1 & 0.36 & 15000 & 9000 $\pm$ 9000 \\ 
 & Outburst 2 & 0.41 $\pm$ 0.13  & 4400 $\pm$ 4100  & 7200 $\pm$ 3400 \\ 
 & Outburst 3 & 0.83 $\pm$ 0.05  & 1700 $\pm$ 400  & 6000 $\pm$ 1500 \\ 
EQ Cha & Quiescent & 0.87 $\pm$ 0.05  & ...  & ...  \\ 
 & Outburst 1 & 0.89 $\pm$ 0.10  & 1700 $\pm$ 530  & 6200 $\pm$ 2000 \\ 
 & Outburst 2 & 0.76 $\pm$ 0.08  & 4700 $\pm$ 1600  & 3500 $\pm$ 1200 \\ 
 & Outburst 3 & 1.0 $\pm$ 1.0  & 360 & 1600 $\pm$ 400 \\ 
HD 35408 & Quiescent & 0.067 $\pm$ 0.002  & ...  & ...  \\ 
HD 287848 & Quiescent & 0.12 $\pm$ 0.00  & ...  & ...  \\ 
2MASS J05250014+0138252 & Quiescent & 0.046 $\pm$ 0.001  & ...  & ...  \\ 
 & Outburst & 0.094 & 51 & 3600 \\ 
\hline
\end{longtable}

\FloatBarrier
\clearpage

\begin{longtable}{llccccc}
\caption{Spectral fitting results with a multi-temperature model}\label{table:sp} \\
\hline\hline
Star & Phase & weighted-$kT$ & $EM$ & $L_{\rm X}$ & $N_H$ & abundance \\ 
 &  & [${\rm keV}$] & [$10^{53}{\rm cm}^{-3}$] & $[10^{30} \mathrm{erg\,s^{-1}}]$ & [$10^{22}{\rm cm}^{-2}$] & [Z$_\odot$] \\ 
\hline
\endfirsthead
\caption{continued.}\\
\hline\hline
Star & Phase & weighted-$kT$ & $EM$ & $L_{\rm X}$ & $N_H$ & abundance \\ 
 &  & [${\rm keV}$] & [$10^{53}{\rm cm}^{-3}$] & $[10^{30} \mathrm{erg\,s^{-1}}]$ & [$10^{22}{\rm cm}^{-2}$] & [Z$_\odot$] \\  
\hline
\endhead
\hline
\endfoot
\multicolumn{7}{@{}l@{}}{\parbox{\linewidth}{\footnotesize
\noindent
\tablefoot{Blank entries indicate cases with insufficient photon statistics or parameters fixed to the quiescent component.} \\
\tablefoottext{a}{
As reported in \citet{2007A&A...461..669R}.
}\\
\tablefoottext{b}{
As reported in \citet{2003A&A...401..543P}.
}\\
\tablefoottext{c}{
As reported in \citet{2010A&A...524A..97L}.
}\\
\tablefoottext{d}{
Previously analyzed by \citet{2023RAA....23j5016R}; the spectral parameters reported therein are adopted. The quoted X-ray luminosity corresponds to the 0.2–8~keV energy band.
}
}}
\endlastfoot
CHX 15b\tablefootmark{a} & Quiescent & $0.75_{-0.04}^{+0.24}$ & $3.7_{-1.1}^{+1.3}$ & $3.5_{-1.2}^{+1.0}$ & $0.24_{-0.14}^{+	0.06}$ & $0.13_{-	0.03}^{+0.04}$ \\
 & Outburst  & $2.4_{-1.0}^{+1.8}$ & $3.4_{-1.3}^{+	1.9}$ & $3.9_{-0.9}^{+1.1}$ & $0.11_{-0.10}^{+0.13}$ & $\leq 0.55$ \\ 
Hen 3-545 & Quiescent & $\geq 0.65$ & $7.6_{-2.1}^{+3.1}$ & $7.3_{-1.4}^{+2.3}$ & $0.33_{-0.05}^{+0.06}$ & $0.15_{-0.03}^{+0.06}$ \\
 & Outburst  & $4.0_{-1.9}^{+2.0}$ & $0.32_{-0.05}^{+0.49}$ & $1.1_{-0.3}^{+0.3}$ & $\leq 0.10$ & $\geq 0.28$ \\
VZ Cha\tablefootmark{a} & Quiescent & $1.0_{-0.1}^{+0.1}$ & $0.89_{-0.15}^{+0.28}$ & $0.79_{-0.12}^{+0.13}$ & $0.17_{-0.04}^{+0.03}$ & $0.05_{-0.01}^{+0.02}$ \\
 & Outburst 1 & $4.2_{-1.4}^{+3.2}$ & $2.2_{-0.5}^{+0.8}$ & $4.4_{-0.6}^{+0.7}$ & $0.52_{-0.14}^{+0.21}$ & $\leq 0.42$ \\
 & Outburst 2 & $4.0_{-0.6}^{+0.8}$ & $2.5_{-0.3}^{+0.4}$ & $4.4_{-0.3}^{+0.4}$ & $0.33_{-0.05}^{+0.05}$ & $0.41_{-0.21}^{+0.35}$ \\
 & Outburst 3 & $3.3_{-1.0}^{+1.9}$ & $2.9_{-0.8}^{+0.9}$ & $0.79_{-0.10}^{+0.10}$ & $0.32_{-0.08}^{+0.09}$ & $\leq 1.0$ \\
V557 Ori & Quiescent & $0.97_{-0.08}^{+0.08}$ & $5.2_{-	0.8}^{+	0.9}$ & $4.1_{-0.5}^{+0.5}$ & $0.06_{-0.02}^{+0.02}$ & $0.03_{-0.01}^{+0.01}$ \\
V569 Ori & Quiescent & $\geq 3.1$ & $33_{-6}^{+5}$ & $40_{-5}^{+4}$ & $0.07_{-0.02}^{+0.02}$ & $0.04_{-0.02}^{+0.03}$ \\
 & Outburst  & $4.0_{-1.4}^{+2.5}$ & $17_{-2}^{+3}$ & $26_{-5}^{+6}$ & $\leq 0.82$ & $\leq 0.82$ \\
V575 Ori & Quiescent & $1.8_{-0.2}^{+0.6}$ & $35_{-11}^{+16}$ & $40_{-10}^{+13}$ & $\leq 0.02$ & $0.13_{-0.07}^{+0.06}$ \\
V984 Ori & Quiescent & $2.5_{-0.4}^{+0.5}$ & $7.0_{-0.7}^{+1.2}$ & $11_{-1}^{+1}$ & $\leq 0.02$ & $0.43_{-0.19}^{+0.23}$ \\
V1044 Ori & Quiescent & $0.71_{-0.06}^{+0.15}$ & $10_{-3}^{+6}$ & $8.3_{-2.3}^{+3.2}$ & $0.18_{-0.07}^{+0.07}$ & $0.09_{-0.02}^{+0.03}$ \\
 & Outburst 1 & $0.78_{-0.24}^{+0.26}$ & $6.4_{-2.7}^{+3.3}$ & $6.1_{-2.5}^{+2.8}$ & ... & ... \\
 & Outburst 2 & $\geq 1.4$ & $2.4_{-0.7}^{+1.0}$ & $3.4_{-1.2}^{+3.8}$ & $\leq 0.82$ & ... \\
V1221 Ori & Quiescent & $1.0_{-0.1}^{+0.0}$ & $10_{-1}^{+2}$ & $8.6_{-0.6}^{+1.1}$ & $\leq 0.02$ & $0.06_{-0.02}^{+0.02}$ \\
V1414 Ori & Quiescent & $1.4_{-0.3}^{+0.5}$ & $55_{-22}^{+49}$ & $54_{-16}^{+27}$ & $0.37_{-0.07}^{+0.08}$ & $0.12_{-0.05}^{+0.13}$ \\
 & Outburst 1 & $12_{-6}^{+6}$ & $98_{-25}^{+26}$ & $220_{-50}^{+50}$ & $0.25_{-0.01}^{+0.01}$ & $0.75_{-0.14}^{+0.15}$ \\
 & Outburst 2 & ... & ... & ... & ... & ... \\
V1417 Ori & Quiescent & $1.1_{-0.8}^{+0.5}$ & $1.3_{-0.7}^{+1.0}$ & $1.4_{-0.9}^{+1.0}$ & $\leq 0.10$ & $0.13_{-0.06}^{+0.09}$ \\
 & Outburst 1 & $0.50_{-0.11}^{+0.25}$ & $51_{-35}^{+179}$ & $23_{-13}^{+52}$ & $0.31_{-0.13}^{+0.15}$ & $\leq 0.82$ \\ 
 & Outburst 2 & $\geq 2.3$ & $9.3_{-4.4}^{+3.6}$ & $14_{-5}^{+5}$ & $\leq 0.82$ & $0.15_{-0.07}^{+0.14}$ \\
V1553 Ori & Quiescent & $\geq 2.0$ & $35_{-17}^{+17}$ & $51_{-19}^{+14}$ & $\leq 0.07$ & $0.05_{-0.05}^{+0.16}$ \\
V1667 Ori & Quiescent & $0.94_{-0.08}^{+0.07}$ & $2.5_{-0.5}^{+0.8}$ & $2.2_{-0.3}^{+0.5}$ & $\leq 0.06$ & $0.06_{-0.02}^{+0.03}$ \\
 & Outburst  & $1.2_{-0.4}^{+1.1}$ & $3.0_{-0.9}^{+4.8}$ & $2.7_{-0.6}^{+2.8}$ & $\leq 0.17$ & $\leq 0.13$ \\
V1740 Ori & Quiescent & $2.2_{-0.7}^{+0.7}$ & $2.2_{-0.6}^{+1.6}$ & $5.3_{-1.1}^{+1.9}$ & $\leq 0.17$ & $1.0_{-0.4}^{+1.4}$ \\
V1760 Ori & Quiescent & $0.71_{-0.03}^{+0.03}$ & $20_{-4}^{+5}$ & $15_{-1}^{+2}$ & $0.10_{-0.03}^{+0.03}$ & $0.08_{-0.02}^{+0.02}$ \\
 & Outburst 1 & $2.0_{-1.2}^{+3.0}$ & $2.6_{-1.4}^{+1.6}$ & $2.7_{-1.0}^{+1.5}$ & $\leq 0.22$ & $\leq 1.7$ \\
 & Outburst 2 & $4.5_{-2.3}^{+26.1}$ & $1.8_{-1.1}^{+1.3}$ & $3.2_{-0.9}^{+1.2}$ & $0.23_{-0.15}^{+0.60}$ & $\leq 2.6$ \\
 & Outburst 3 & $3.9_{-0.9}^{+1.1}$ & $3.9_{-0.9}^{+1.2}$ & $7.0_{-0.9}^{+1.0}$ & $0.06_{-0.05}^{+0.05}$ & $0.56_{-0.41}^{+0.75}$ \\
V1878 Ori & Quiescent & $1.4_{-0.3}^{+0.7}$ & $21_{-6}^{+5}$ & $19_{-4}^{+4}$ & $0.03_{-0.02}^{+0.02}$ & $0.11_{-0.02}^{+0.06}$ \\ 
 & Outburst 1 & $2.9_{-1.8}^{+2.2}$ & $0.81_{-0.51}^{+2.39}$ & $1.8_{-0.6}^{+0.7}$ & $\leq 0.14$ & $\leq 4.8$ \\
 & Outburst 2 & $2.5_{-0.8}^{+1.3}$ & $4.1_{-0.8}^{+0.9}$ & $5.3_{-0.8}^{+1.0}$ & $\leq 0.04$ & $\leq 0.52$ \\ 
V2087 Ori & Quiescent & $1.7_{-0.1}^{+2.1}$ & $9.9_{-4.7}^{+2.7}$ & $15_{-6}^{+4}$ & $\leq 0.01$ & $0.36_{-0.09}^{+0.08}$ \\ 
V2098 Ori & Quiescent & $0.67_{-0.05}^{+0.11}$ & $7.8_{-2.5}^{+3.9}$ & $5.6_{-1.2}^{+1.8}$ & $0.10_{-0.05}^{+0.04}$ & $0.06_{-0.02}^{+0.03}$ \\ 
V2108 Ori & Outburst 1 & $5.4_{-1.8}^{+3.6}$ & $17_{-4}^{+6}$ & $32_{-4}^{+6}$ & $2.7_{-0.5}^{+0.6}$ & $0.37_{-0.27}^{+0.35}$ \\ 
 & Outburst 2 & $5.7_{-1.6}^{+3.0}$ & $11_{-2}^{+3}$ & $22_{-3}^{+3}$ & $2.5_{-0.4}^{+0.4}$ & $0.22_{-0.19}^{+0.20}$ \\
AR Ori & Quiescent & $1.0_{-0.1}^{+0.1}$ & $5.9_{-	1.1}^{+1.8}$ & $5.7_{-0.9}^{+1.1}$ & $\leq 0.06$ & $0.08_{-0.03}^{+0.04}$ \\ 
 & Outburst  & ... & ... & ... & ... & ... \\ 
PR Ori & Quiescent & $\geq 1.2$ & $9.4_{-3.9}^{+3.9}$ & $12_{-4}^{+4}$ & $0.03_{-0.02}^{+0.03}$ & $0.21_{-0.06}^{+0.17}$ \\ 
WW Ori & Quiescent & $1.1_{-0.1}^{+0.1}$ & $9.0_{-0.9}^{+1.1}$ & $8.2_{-0.6}^{+0.7}$ & $0.18_{-0.02}^{+0.02}$ & $0.06_{-0.01}^{+0.02}$ \\
YZ Ori & Quiescent & $0.44_{-0.02}^{+0.02}$ & $16_{-5}^{+11}$ & $14_{-3}^{+7}$ & $\leq 0.09$ & $0.16_{-0.05}^{+0.07}$ \\ 
Brun 213 & Quiescent & $1.1_{-0.1}^{+0.2}$ & $7.5_{-1.5}^{+2.4}$ & $9.9_{-1.6}^{+2.2}$ & $0.06_{-0.04}^{+0.04}$ & $0.30_{-0.11}^{+0.20}$ \\
 & Outburst  & $3.1_{-1.7}^{+4.6}$ & $5.4_{-2.2}^{+1.3}$ & $7.2_{-2.1}^{+3.5}$ & $\leq 0.82$ & $\leq 1.8$ \\
Brun 302 & Quiescent & $0.70_{-0.03}^{+0.04}$ & $33_{-7}^{+11}$ & $28_{-5}^{+8}$ & $0.14_{-0.04}^{+0.05}$ & $0.10_{-0.02}^{+0.03}$ \\
Brun 311 & Quiescent & $1.6_{-0.2}^{+0.6}$ & $49_{-13}^{+21}$ & $62_{-13}^{+18}$ & $0.30_{-0.04}^{+0.07}$ & $0.21_{-0.07}^{+0.12}$ \\ 
 & Outburst  & $3.4_{-0.9}^{+1.2}$ & $21_{-4}^{+6}$ & $31_{-4}^{+4}$ & $0.20_{-0.05}^{+0.07}$ & $\leq 0.50$ \\
Brun 394 & Quiescent & $\geq 1.2$ & $7.7_{-1.6}^{+2.0}$ & $8.4_{-2.2}^{+3.1}$ & $\leq 0.04$ & $0.08_{-0.03}^{+0.05}$ \\
 & Outburst 1 & $4.1_{-1.2}^{+2.6}$ & $6.0_{-0.9}^{+1.2}$ & $12_{-2}^{+2}$ & $\leq 0.04$ & $\leq 1.4$ \\
 & Outburst 2 & $2.2_{-0.6}^{+3.7}$ & $0.95_{-0.78}^{+5.02}$ & $1.6_{-0.8}^{+1.6}$ & $0.00_{-0.00}^{+0.00}$ & $\leq 3.1$ \\
Brun 656 & Quiescent & $1.4_{-0.2}^{+0.4}$ & $110_{-20}^{+20}$ & $110_{-10}^{+20}$ & $0.34_{-0.04}^{+0.03}$ & $0.12_{-0.02}^{+0.03}$ \\
 & Outburst 1 & $3.1_{-1.4}^{+3.1}$ & $9.7_{-3.3}^{+6.3}$ & $14_{-3}^{+4}$ & $0.25_{-0.10}^{+0.16}$ & $\leq 0.79$ \\
 & Outburst 2 & $3.7_{-0.6}^{+0.9}$ & $24_{-4}^{+4}$ & $40_{-4}^{+4}$ & $0.22_{-0.05}^{+0.05}$ & $0.38_{-0.27}^{+0.33}$ \\
Brun 684 & Quiescent & $1.4_{-0.2}^{+0.3}$ & $7.5_{-1.7}^{+3.0}$ & $12_{-6}^{+5}$ & $\leq 0.08$ & $0.44_{-0.17}^{+0.25}$ \\
BD-06 1236 & Quiescent & $0.86_{-0.15}^{+0.14}$ & $1.0_{-0.2}^{+1.0}$ & $1.5_{-0.3}^{+0.6}$ & $\leq 0.12$ & $0.30_{-0.15}^{+0.12}$ \\ 
 & Outburst 1 & ... & ... & ... & ... & ... \\
 & Outburst 2 & $1.8_{-0.3}^{+0.5}$ & $3.2_{-0.5}^{+0.5}$ & $3.6_{-0.3}^{+0.3}$ & $\leq 0.02$ & $0.13_{-0.09}^{+0.16}$ \\
HD 36959 & Quiescent & $1.1_{-0.1}^{+0.1}$ & $6.7_{-1.2}^{+2.2}$ & $6.2_{-1.1}^{+1.5}$ & $\leq 0.06$ & $0.07_{-0.03}^{+0.04}$ \\
HD 36960 & Quiescent & $\geq 0.88$ & $29_{-	5}^{+11}$ & $32_{-5}^{+5}$ & $\leq 0.03$ & $0.18_{-0.06}^{+0.09}$ \\
HRC 132 & Quiescent & $0.85_{-0.07}^{+0.09}$ & $51_{-14}^{+32}$ & $40_{-7}^{+15}$ & $0.10_{-0.05}^{+0.06}$ & $0.08_{-0.03}^{+0.05}$ \\
Parenago 1578 & Quiescent & $2.0_{-0.6}^{+2.5}$ & $160_{-50}^{+50}$ & $160_{-40}^{+30}$ & $0.16_{-0.04}^{+0.03}$ & $0.07_{-0.02}^{+0.05}$ \\
UCAC4 427-010050 & Quiescent & $0.65_{-0.03}^{+0.03}$ & $1.3_{-0.1}^{+0.2}$ & $1.2_{-0.1}^{+0.1}$ & $\leq 0.82$ & $0.15_{-0.03}^{+0.04}$ \\ 
 & Outburst  & $0.76_{-0.56}^{+0.29}$ & $0.29_{-0.13}^{+0.15}$ & $0.33_{-0.15}^{+0.16}$ & ... & ... \\ 
2MASS J05343190-0526368 & Quiescent & $0.94_{-0.17}^{+0.18}$ & $0.97_{-0.20}^{+0.59}$ & $0.78_{-	0.07}^{+0.31}$ & $\leq 0.08$ & $0.03_{-0.03}^{+0.03}$ \\ 
 & Outburst  & ... & ... & ... & ... & ... \\ 
2MASS J05353534-0511114 & Quiescent & $\geq 0.99$ & $23_{-8}^{+8}$ & $23_{-6}^{+6}$ & $0.07_{-0.05}^{+0.04}$ & $0.09_{-0.03}^{+0.10}$ \\ 
 & Outburst 1 & $\geq 1.6$ & $4.7_{-2.2}^{+2.9}$ & $8.5_{-4.6}^{+10.8}$ & ... & ... \\ 
 & Outburst 2 & $0.98_{-0.25}^{+0.39}$ & $63_{-30}^{+63}$ & $53_{-21}^{+58}$ & $0.25_{-0.16}^{+0.42}$ & $\leq 0.25$ \\ 
ASR 106\tablefootmark{b} & Quiescent & $2.3_{-0.4}^{+0.7}$ & $10_{-3}^{+4}$ & $13_{-3}^{+4}$ & $1.9_{-0.4}^{+0.4}$ & $\leq 0.62$ \\
 & Outburst  & $7.4_{-1.2}^{+1.6}$ & $51_{-5}^{+5}$ & $110_{-10}^{+10}$ & $1.8_{-0.2}^{+0.2}$ & $0.52_{-0.19}^{+0.22}$ \\
EM*~LkHA~270\tablefootmark{b} & Quiescent & $0.45_{-0.07}^{+0.12}$ & $130_{-50}^{+110}$ & $65_{-25}^{+55}$ & $0.83_{-0.12}^{+0.14}$ & $0.03_{-0.02}^{+0.03}$ \\
 & Outburst  & $1.2_{-0.3}^{+4.2}$ & $4.4_{-3.3}^{+6.2}$ & $6.3_{-3.8}^{+8.6}$ & $0.96_{-0.65}^{+0.57}$ & $\geq 0$ \\
Gaia DR3 121032079618050560 & Quiescent & $\geq 1.3$ & $5.6_{-2.6}^{+11.8}$ & $6.8_{-2.7}^{+6.0}$ & $0.09_{-0.06}^{+0.12}$ & $0.21_{-0.16}^{+0.35}$ \\
 & Outburst  & $1.0_{-0.2}^{+0.2}$ & $1.8_{-0.7}^{+0.7}$ & $1.9_{-0.4}^{+0.4}$ & $\leq 0.82$ & $0.12_{-0.08}^{+0.15}$ \\
EI Cha\tablefootmark{c} & Quiescent & $0.59_{-0.02}^{+0.02}$ & $1.4_{-0.2}^{+0.3}$ & $1.2_{-0.1}^{+0.2}$ & $0.05_{-0.03}^{+0.03}$ & $0.12_{-0.02}^{+0.02}$ \\
 & Outburst 1 & $\geq 1.6$ & $0.069_{-0.050}^{+0.040}$ & $0.15_{-0.09}^{+0.12}$ & $\leq 13$ & ... \\ 
 & Outburst 2 & $\geq 2.4$ & $1.3_{-0.5}^{+0.3}$ & $1.8_{-0.3}^{+0.4}$ & $\leq 0.82$ & $0.10_{-0.07}^{+0.15}$ \\ 
EL Cha\tablefootmark{c} & Quiescent & $0.67_{-0.04}^{+0.04}$ & $0.36_{-0.05}^{+0.05}$ & $0.32_{-0.04}^{+0.03}$ & $\leq 0.82$ & $0.11_{-0.03}^{+0.04}$ \\
 & Outburst 1 & $0.90_{-0.33}^{+0.43}$ & $0.27_{-0.18}^{+0.57}$ & $0.25_{-0.10}^{+0.28}$ & $\leq 0.23$ & $\leq 0.35$ \\
 & Outburst 2 & $0.97_{-0.46}^{+5.14}$ & $0.39_{-0.16}^{+0.17}$ & $0.40_{-0.16}^{+0.16}$ & $\leq 0.82$ & $0.11_{-0.03}^{+0.04}$ \\ 
EM Cha\tablefootmark{c} & Quiescent & $\geq 0.71$ & $2.6_{-0.4}^{+0.5}$ & $2.3_{-0.3}^{+0.3}$ & $0.03_{-0.02}^{+0.02}$ & $0.12_{-0.02}^{+0.04}$ \\
EN Cha\tablefootmark{c} & Quiescent & $0.83_{-0.14}^{+0.13}$ & $0.45_{-0.40}^{+0.23}$ & $0.56_{-0.19}^{+0.58}$ & $0.42_{-0.17}^{+0.37}$ & $\geq 0.08$ \\
 & Outburst  & $2.3_{-0.3}^{+0.4}$ & $0.73_{-0.14}^{+0.18}$ & $0.98_{-0.11}^{+0.12}$ & $0.27_{-0.06}^{+0.08}$ & $0.30_{-0.18}^{+0.27}$ \\
EO Cha\tablefootmark{c} & Quiescent & $0.73_{-0.04}^{+0.04}$ & $0.63_{-0.08}^{+0.18}$ & $0.62_{-0.05}^{+0.12}$ & $\leq 0.04$ & $0.15_{-0.04}^{+0.06}$ \\
EP Cha\tablefootmark{c} & Quiescent & $\geq 0.46$ & $9.4_{-5.6}^{+4.8}$ & $8.0_{-3.2}^{+3.2}$ & $0.17_{-0.09}^{+0.05}$ & $0.15_{-0.03}^{+0.20}$ \\ 
 & Outburst 1 & $2.3_{-0.6}^{+0.8}$ & $0.39_{-0.23}^{+0.37}$ & $0.71_{-0.17}^{+0.22}$ & $\leq 0.15$ & $1.0_{-0.8}^{+2.1}$ \\ 
 & Outburst 2 & $2.1_{-0.9}^{+1.4}$ & $0.64_{-0.31}^{+0.67}$ & $0.76_{-0.20}^{+0.16}$ & $\leq 0.24$ & $\leq 1.0$ \\
 & Outburst 3 & $2.7_{-0.6}^{+0.8}$ & $0.87_{-0.21}^{+0.22}$ & $1.3_{-0.2}^{+0.2}$ & $\leq 0.82$ & $0.44_{-0.38}^{+0.60}$ \\
EQ Cha\tablefootmark{c} & Quiescent & $0.80_{-0.06}^{+0.07}$ & $1.6_{-0.3}^{+0.4}$ & $1.6_{-0.2}^{+0.2}$ & $\leq 0.82$ & $0.16_{-0.07}^{+0.12}$ \\
 & Outburst 1 & $\geq 1.5$ & $0.23_{-0.18}^{+0.31}$ & $0.57_{-0.35}^{+0.62}$ & $\leq 13$ & ... \\
 & Outburst 2 & $\geq 0.99$ & $0.48_{-0.39}^{+0.53}$ & $0.69_{-0.32}^{+0.89}$ & $\leq 0.82$ & ... \\
 & Outburst 3 & ... & ... & ... & ... & ... \\
HD 35408 & Quiescent & $\geq 1.5$ & $1.4_{-0.6}^{+1.7}$ & $2.6_{-0.7}^{+1.2}$ & $\leq 0.11$ & $0.54_{-0.37}^{+1.18}$ \\
HD 287848\tablefootmark{d} & Quiescent & $1.0_{-0.1}^{+0.1}$ & $3.6_{-0.5}^{+0.6}$ & $6.0_{-1.0}^{+1.0}$ & $\leq 0.02$ & $0.39_{-0.09}^{+0.09}$ \\
2MASS J05250014+0138252 & Quiescent & $0.88_{-0.05}^{+0.05}$ & $2.3_{-0.4}^{+0.5}$ & $2.2_{-0.2}^{+0.4}$ & $\leq 0.04$ & $0.09_{-0.02}^{+0.03}$ \\ 
 & Outburst  & $3.0_{-1.8}^{+7.1}$ & $2.5_{-1.7}^{+5.4}$ & $3.7_{-0.7}^{+3.0}$ & $\leq 0.47$ & ... \\
\hline
\end{longtable}

\clearpage
\end{appendix}

\end{document}